\pdfoutput=1
\documentclass[a4paper,12pt]{article}

\usepackage{fix-cm}
\usepackage{amsfonts}
\usepackage{mathrsfs}
\usepackage{amssymb}
\usepackage{amsmath}
\usepackage{shuffle}
\usepackage{framed}
\usepackage[medium]{titlesec}
\usepackage{bm}
\usepackage{cite}
\usepackage[normalem]{ulem}
\usepackage{extarrows}
\usepackage{slashed}
\usepackage{isodateo}
\usepackage{graphicx}
\usepackage{array}
\usepackage[dvipsnames]{xcolor}
\usepackage[hmargin=.7in,vmargin=1.1in]{geometry}
\usepackage{indentfirst}
\usepackage{booktabs}
\usepackage{bbm}
\usepackage{float}
\usepackage{mdframed}
\usepackage[bookmarksnumbered=true,bookmarksopen=true]{hyperref}
\hypersetup{colorlinks=true,linkcolor=NavyBlue,citecolor=Periwinkle,urlcolor=Periwinkle}

\newcommand{\ma}{\mathcal}
\newcommand{\dd}{\mathrm d}
\newcommand{\ii}{\mathrm i}

\newcommand{\mpl}{M_{\rm Pl}}
\newcommand{\mb}{\mathbf}
\newcommand{\tr}{\operatorname{Tr}}
\newcommand{\intp}{\int\!\frac{\dd^3p}{(2\pi)^3}}

\newcommand{\fnemail}[1]{\footnote{Email: \href{mailto:#1}{\nolinkurl{#1}}}}

\newmdenv[backgroundcolor=gray!15,
          skipabove=0pt,
          skipbelow=5pt,
          leftmargin=0pt,
          rightmargin=0pt,
          innertopmargin=-5pt,
          innerbottommargin=7pt,
          innerleftmargin=2pt,
          innerrightmargin=2pt,
          splittopskip=0pt,
          splitbottomskip=0pt,
          linewidth=0pt,
          nobreak=true]{keyeqn}

\titleformat{\section}
{\normalfont\fontsize{15}{20}\bfseries}{\thesection}{1em}{}

\begin{document}

\title{\Large\textbf{Heavy-particle production during inflation and its gravitational-wave signal}}

\author{Qi Chen$^{\,a\,}$\fnemail{chenqi25@ucas.ac.cn}~~~~~ and ~~~~~Yuan Yin$^{\,b\,}$\fnemail{yinyuan@kias.re.kr}\\
\normalsize{\rule{0pt}{6mm}\emph{$^a\,$Kavli Institute for Theoretical Sciences, University of Chinese Academy of Sciences,}}\\
\normalsize{\emph{Beijing 100190, China}}\\
$^b\,$\normalsize{\emph{School of Physics, Korea Institute for Advanced Study, Seoul 02455, Korea}}
}

\date{}
\maketitle

\vspace{5mm}

\begin{abstract}
\vspace{10mm}
We study a mechanism for producing heavy particles during inflation through a transient quadratic $U(1)$-breaking interaction of a complex scalar coupled derivatively to the inflaton. The rolling inflaton induces an effective chemical potential, while sufficiently strong symmetry breaking creates a tachyonic instability that amplifies the scalar fluctuations. The finite duration of the instability limits their growth, allowing efficient production of particles heavier than the Hubble scale in regimes where homogeneous backreaction remains small. The amplified fluctuations source a peaked gravitational-wave spectrum through their anisotropic stress. We calculate this spectrum by following the full time evolution of the source. For suitable parameters and production times, our idealized forecasts suggest that the signal could be observable with LISA, subject to the assumed switching profile and reheating history.
\end{abstract}

\newpage
\tableofcontents
\newpage

\section{Introduction}

Inflation resolves the horizon problem and generates the primordial perturbations that seed large-scale structure~\cite{Guth:1980zm,Starobinsky:1979ty,Linde:1981mu,Albrecht:1982wi}. It may also provide the setting in which the matter--antimatter asymmetry originated~\cite{Affleck:1984fy,Mohapatra:2021aig,Hertzberg:2013mba,Lozanov:2014zfa,Yamada:2015xyr,Bamba:2016vjs,Bamba:2018bwl,Charng:2008ke,Wu:2019ohx,Lin:2020lmr,Barrie:2020hiu,An:2024zfi}. Since producing this asymmetry requires baryon-number violation, inflationary baryogenesis motivates theories with broken $U(1)$ charge. A quadratic breaking term for a complex scalar provides a minimal realization.

Previous studies~\cite{Bodas:2020yho,Bodas:2024hih,Bodas:2025wuk} showed that explicit breaking renders physical the chemical potential induced by a dimension-five inflaton-current interaction; without breaking, a time-dependent field redefinition removes it. The same field content can therefore connect particle production with cosmological-collider physics. We focus here on the finite-duration production history and its tensor source.

Inflationary particle production also provides a possible origin for superheavy relics. Wimpzilla candidates~\cite{Kolb:1998ki}, for example, can be heavier than the inflationary Hubble scale, where purely gravitational production is Boltzmann suppressed. Proposed non-adiabatic alternatives include phase transitions during inflation~\cite{An:2022toi}, a sharp transition at the end of inflation~\cite{Li:2019ves,Li:2020xwr}, axion backgrounds~\cite{Adshead:2019aac}, and inflaton oscillations~\cite{Kofman:1997yn,Ema:2015dka,Ema:2016hlw,Ema:2018ucl}; see Ref.~\cite{Kolb:2023ydq} for a review. A finite interval of $U(1)$ breaking provides another production mechanism, whose efficiency we quantify below.

The model combines the breaking operator $\chi^2+\chi^{*2}$ with the derivative coupling $\partial_\mu\phi\,\ma J^\mu/\Lambda$. It produces finite-duration tachyonic amplification for the $m/H=\mu/H=2$ benchmarks with $A/H=4$ and $4.5$. The finite active interval limits the growth, while the chemical potential controls its exponent, the rotation of the eigenvectors, and the charge asymmetry. The stability criterion is unchanged in a rotating field basis, where the breaking phase oscillates at $2\mu$.

The produced $\chi$ fluctuations carry anisotropic stress and source tensor perturbations~\cite{Cook:2011hg}. We express the source in terms of the physical spatial stress and fix its normalization through the inverse helicity projector. Its unequal-time correlator includes both real components and the anomalous contractions induced by $U(1)$ breaking. Our primary tensor quantity is the difference between the $A\neq0$ and $A=0$ stress correlators evaluated with the same regulator. We integrate the resulting finite-time kernel using the mode functions evolved through the smooth production interval.

Current and planned observations probe stochastic gravitational waves over many decades in frequency. Ground-based interferometers operate in the Hz--kHz range~\cite{LIGOScientific:2022sts}, space missions target the mHz band~\cite{Audley:2017drz,LISA:2024mow,Luo:2015ght,Guo:2018npi}, pulsar-timing arrays probe nHz scales~\cite{NANOGrav:2023hde,EPTA:2023sfo}, and CMB polarization constrains the lowest frequencies~\cite{CMB-S4:2016ple}. Mapping an inflationary production scale to an observed frequency requires the production time, its relation to the CMB pivot, the expansion during reheating, and the subsequent cosmological history. The $A/H=4$ case provides a numerical check of the finite-time tensor calculation, with a sourced spectrum far below the standard vacuum contribution. We also study two benchmark points with a shorter production interval that satisfy the time-dependent backreaction bounds. For a specified production time and reheating history, their spectra peak in the LISA band.

Section~\ref{sec:The_Model} introduces the model and derives the local stability criterion. Section~\ref{sec:particleproduction} analyzes smooth finite-duration production and the spectator constraints. Section~\ref{sec:GWPowerspectrum} presents the tensor calculation, its transfer to the present day, and two LISA-band benchmark points. Section~\ref{sec:SecofCC} discusses the relation to cosmological-collider results, and Section~\ref{sec:conclusions} summarizes our conclusions. Numerical details are collected in the appendices.

\section{Model and local stability}\label{sec:The_Model}

Production of a heavy spectator during inflation is exponentially suppressed unless the background evolution violates adiabaticity. For a complex scalar, a chemical potential alone can be removed by a field redefinition, whereas quadratic $U(1)$ breaking fixes the relative phase and couples the two real components.

We use the $(-+++)$ signature and define
\begin{align}
\ma J^\mu\equiv-\ii\left(\chi\partial^\mu\chi^*-\chi^*\partial^\mu\chi\right),
\qquad
\mu\equiv\frac{\dot\phi_0}{\Lambda}.
\end{align}
The covariant spectator action is
\begin{keyeqn}
\begin{align}
S_\chi=\int\dd^4x\sqrt{-g}\bigg[
&-g^{\mu\nu}\partial_\mu\chi^*\partial_\nu\chi
-m^2|\chi|^2
-\frac{\ii}{\Lambda}\partial_\mu\phi
\left(\chi\partial^\mu\chi^*-\chi^*\partial^\mu\chi\right)
\nonumber\\
&-\frac{A^2}{2}\,w[\sigma(x)]
\left(\chi^2+\chi^{*2}\right)
\bigg].
\label{eq:lagcorvariant}
\end{align}
\end{keyeqn}
Here $w[\sigma]$ is a scalar switching profile generated by a trigger field. For a homogeneous FRW background,
\begin{align}
\ma L_{\rm FRW}=a^3\bigg[
|\dot\chi|^2-a^{-2}|\nabla\chi|^2-m^2|\chi|^2
+\ii\mu(\chi\dot\chi^*-\chi^*\dot\chi)
-\frac{A^2}{2}w(t)(\chi^2+\chi^{*2})
\bigg].
\label{eq:lagchemical}
\end{align}
Writing $\chi=(\chi_1+\ii\chi_2)/\sqrt2$, the two non-derivative plateau masses are
\begin{align}
m_1^2=m^2+A^2,\qquad m_2^2=m^2-A^2.
\end{align}
Quadratic terms of this form arise, for example, as supersymmetric $b$-terms~\cite{Martin:1997ns}.
A constant rephasing $\chi\to\ii\chi$ reverses the sign of the breaking term and exchanges the labels $1$ and $2$.

The time-dependent rotation
\begin{align}
\chi=e^{-\ii\Theta(t)}\widetilde\chi,\qquad \dot\Theta=\mu,
\end{align}
removes the derivative interaction. For constant $\mu$ it gives
\begin{keyeqn}
\begin{align}
\ma L_{\rm FRW}=a^3\bigg[
&|\dot{\widetilde\chi}|^2-a^{-2}|\nabla\widetilde\chi|^2
-(m^2+\mu^2)|\widetilde\chi|^2
\nonumber\\
&-\frac{A^2}{2}w(t)
\left(e^{-2\ii\Theta}\widetilde\chi^2
+e^{2\ii\Theta}\widetilde\chi^{*2}\right)
\bigg].
\label{eq:lagredef}
\end{align}
\end{keyeqn}
The breaking phase therefore rotates at $2\mu$. In exact de Sitter,
\begin{align}
a=-\frac{1}{H\tau},
\qquad
\Theta=-\frac{\mu}{H}\ln(-H\tau)
\end{align}
up to a constant. The coupled mode equations are
\begin{align}
\widetilde\chi_{\mb k}''-\frac{2}{\tau}\widetilde\chi_{\mb k}'
+\left[k^2+\frac{m^2+\mu^2}{H^2\tau^2}\right]\widetilde\chi_{\mb k}
+\frac{A^2w}{H^2\tau^2}(-H\tau)^{-2\ii\mu/H}
\widetilde\chi^*_{-\mb k}&=0,
\label{eq:eom1}\\
\widetilde\chi^*_{-\mb k}{}''-\frac{2}{\tau}\widetilde\chi^*_{-\mb k}{}'
+\left[k^2+\frac{m^2+\mu^2}{H^2\tau^2}\right]\widetilde\chi^*_{-\mb k}
+\frac{A^2w}{H^2\tau^2}(-H\tau)^{2\ii\mu/H}
\widetilde\chi_{\mb k}&=0.
\label{eq:eom2}
\end{align}

\paragraph{Local stability.}
Holding the background quantities fixed on time scales short compared with $H^{-1}$ and writing $q=k/a$, the two normal-mode frequencies on the plateau are
\begin{align}
\omega_\pm^2(q)=q^2+m^2+2\mu^2
\pm\sqrt{A^4+4\mu^2(q^2+m^2+\mu^2)}.
\label{eq:dispersion}
\end{align}
Their product is basis independent,
\begin{align}
\omega_+^2\omega_-^2=(q^2+m^2)^2-A^4,
\qquad
\omega_-^2<0\ \Longleftrightarrow\ q^2<A^2-m^2.
\label{eq:dispersionproduct}
\end{align}
The stability boundary is $A=m$, independent of $\mu$. The chemical potential changes the unstable eigenvalue and the charge composition but cannot stabilize an $A>m$ plateau. In the rotating basis, the oscillating phase in Eq.~\eqref{eq:lagredef} describes the same tachyonic dynamics.

\paragraph{Smooth switching.}
We model the finite production interval with smooth turn-on and turn-off transitions,
\begin{align}
w(N)&=s(N-N_i)s(N_*-N),\nonumber\\
s(x)&=\frac12\left[1+\tanh\left(\frac{x}{\sigma}\right)\right],
\qquad
\sigma=\frac{\Delta N_{10-90}}{2\,\operatorname{arctanh}(0.8)},
\label{eq:smoothwindow}
\end{align}
where $N=\ln(a/a_*)$. The fiducial profile has
\begin{align}
N_i=-\ln100,\qquad N_*=0,\qquad \Delta N_{10-90}=0.2.
\end{align}
Equivalently, the turn-on and turn-off midpoints correspond to $H\tau_i=-100$ and $H\tau_*=-1$. We evolve the modes through the smooth tail beyond the turn-off midpoint $N_*$.

One possible microscopic trigger is
\begin{align}
V(\phi,X,\sigma)=V_{\rm inf}(\phi)
+\frac{\lambda_\sigma}{4}(\sigma^2-v^2)^2
+\frac12(M^2-c^2\phi^2)X^2
+\frac{\lambda_X}{4}X^4
+\frac12g^2X^2\sigma^2.
\label{eq:TheTriggerSector}
\end{align}
This potential can make the $\sigma$ expectation value nonzero only transiently. We prescribe $w(N)$ in the calculation below; obtaining it from this potential would require solving the trigger dynamics. The energy transferred through the term proportional to $w_{,N}$ consequently depends on the trigger model.

\begin{figure}[t]
\centering
\includegraphics[width=0.98\textwidth]{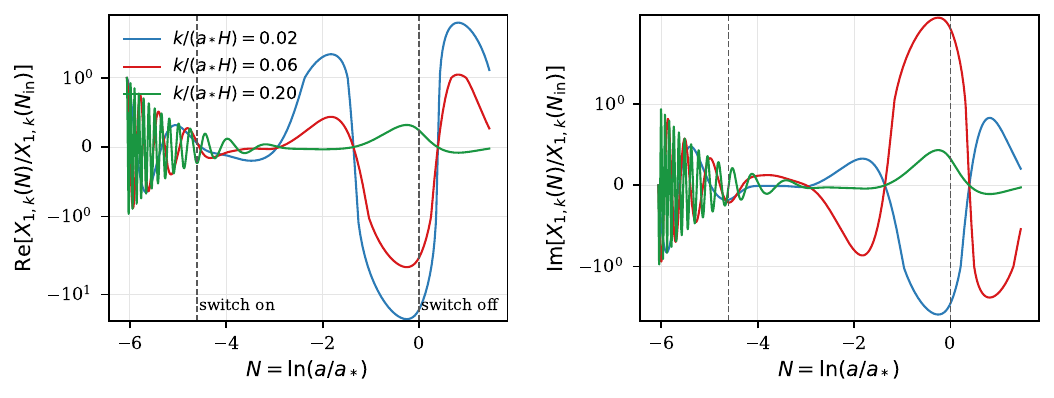}
\caption{Evolution of the real and imaginary parts of the first column of the mode matrix in the physical-field basis, for $m/H=\mu/H=2$, $A/H=4$, $H\tau_i=-100$, $H\tau_*=-1$, and $\Delta N_{10-90}=0.2$. The curves show three values of $k/(a_*H)$; dashed lines mark the turn-on and turn-off midpoints. For visual comparison, the mode functions are normalized to their initial amplitude.}
\label{fig:SolutionofEOM}
\end{figure}

Figure~\ref{fig:SolutionofEOM} shows the first column of the mode matrix in the physical-field basis, evolved from exact Bunch--Davies initial data through the smooth window. Rapid subhorizon oscillations are followed by finite amplification while the breaking term is active. Once $w$ has become negligible, both components again evolve according to the unbroken theory.

\section{Particle production and backreaction}\label{sec:particleproduction}

We evolve both real degrees of freedom through the finite production interval and define particles only after the breaking term has switched off. Introduce
\begin{align}
\widetilde\chi=\frac{X_1+\ii X_2}{\sqrt2},
\qquad
\mb Y=\sqrt H\,a^{3/2}(X_1,X_2)^T,
\qquad
\kappa=\frac{k}{a_*H}.
\end{align}
For the canonical field $u=a\widetilde\chi$, we write
$u=(u_++\ii u_-)/\sqrt2$. Defining
$\theta(\tau)=(2\mu/H)\ln(-H\tau)$, we obtain the component equations
\begin{align}
u_+''+\left[k^2+\frac{m^2+\mu^2}{H^2\tau^2}-\frac{2}{\tau^2}
+\frac{A^2w}{H^2\tau^2}\cos\theta\right]u_+
-\frac{A^2w}{H^2\tau^2}\sin\theta\,u_-&=0,
\label{eq:EomPlus}\\
u_-''+\left[k^2+\frac{m^2+\mu^2}{H^2\tau^2}-\frac{2}{\tau^2}
-\frac{A^2w}{H^2\tau^2}\cos\theta\right]u_-
-\frac{A^2w}{H^2\tau^2}\sin\theta\,u_+&=0.
\label{eq:EomMinus}
\end{align}
We evolve this coupled system with a $2\times2$ complex mode matrix
$\mb F_\kappa$ for the rescaled field $\mb Y$. Its columns propagate the two
independent Bunch--Davies initial modes,
\begin{align}
\frac{\dd^2\mb F_\kappa}{\dd N^2}
+\mb K_\kappa(N)\mb F_\kappa=0,
\label{eq:matrixmode}
\end{align}
with
\begin{align}
\mb K_\kappa(N)=&
\left[\kappa^2e^{-2N}
+\frac{m^2+\mu^2}{H^2}-\frac94\right]\mb 1
+\frac{A^2}{H^2}w(N)
\mb R\!\left(\frac{\mu N}{H}\right)
\mb\sigma_3
\mb R^T\!\left(\frac{\mu N}{H}\right),
\label{eq:rotatedstiffness}\\
\mb R(\vartheta)&=
\begin{pmatrix}
\cos\vartheta&-\sin\vartheta\\
\sin\vartheta&\cos\vartheta
\end{pmatrix},
\qquad
\mb\sigma_3=\operatorname{diag}(1,-1).
\label{eq:rotationmatrix}
\end{align}
The derivative-coupled and rotated representations give identical late-time observables.

No basis-independent instantaneous particle number exists while $\omega_-^2<0$. Once $w$ is negligible, we project the mode matrix onto the exact $A=0$ de Sitter solution evolved from the same Bunch--Davies state. If $\mb\alpha_k$ and $\mb\beta_k$ denote the resulting scattering matrices, we define
\begin{align}
\mb{\mathsf N}^{\rm ind}_k=\mb\beta_k^*\mb\beta_k^T,
\qquad
n_k^{\rm ind}=\tr\mb{\mathsf N}^{\rm ind}_k.
\label{eq:occupationmatrix}
\end{align}
The occupation number $n_k^{\rm ind}$ is defined relative to the exact $A=0$ solution, with the trace counting both real degrees of freedom. Appendix~\ref{sec:Bogliubov} gives the projection and numerical checks.

For each mode, we define the basis-independent tachyonic growth exponent
\begin{align}
\ma I_{\rm tach}(k)=
\int_{\omega_-^2(k,N)<0}\dd N\,
\frac{\sqrt{-\omega_-^2(k,N)}}{H}.
\label{eq:tachexposure}
\end{align}
For $m/H=\mu/H=2$, more than $99.9\%$ of $\rho_{\rm occ}^{\rm ind}$ comes from modes that enter the local tachyonic band, for both $A/H=4$ and $4.5$. Fits of $\ln(1+n_k^{\rm ind})$ to $2\ma I_{\rm tach}$ have slopes $1.06$--$1.09$ and $R^2>0.999$. The calculation with the background held fixed in Appendix~\ref{sec:Bogliubov} likewise exhibits exponential growth only when $\omega_-^2<0$. These tests identify finite-duration tachyonic growth as the dominant production mechanism.

\begin{figure}[t]
\centering
\includegraphics[width=0.98\textwidth]{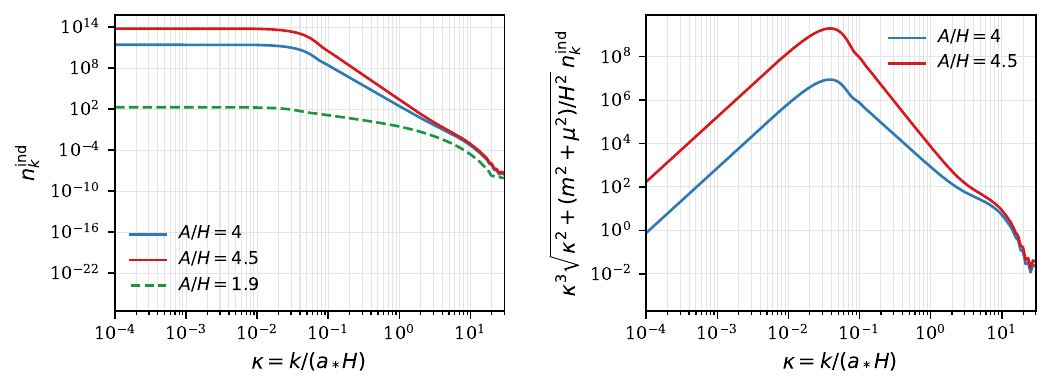}
\caption{Occupation spectra after the breaking interaction has switched off, for $m/H=\mu/H=2$, $H\tau_i=-100$, $H\tau_*=-1$, and the fiducial tanh profile with $\Delta N_{10-90}=0.2$. Left: the occupation $n_k^{\rm ind}$ relative to the $A=0$ solution for $A/H=4$, $4.5$, and the stable case $A/H=1.9$. Right: the occupation-energy weight per logarithmic momentum interval entering Eq.~\eqref{eq:rhop_def}, for $A/H=4$ and $4.5$. The occupation numbers in both panels are evaluated once $w$ has become negligible.}
\label{fig:TheNumberDensity}
\end{figure}

For comparisons among switching profiles, we define the energy-weighted occupation integral
\begin{keyeqn}
\begin{align}
\rho_{\rm occ}^{\rm ind}(a_*)
=\frac{1}{2\pi^2a_*^4}
\int\dd k\,k^2\sqrt{k^2+a_*^2(m^2+\mu^2)}
\,n_k^{\rm ind}.
\label{eq:rhop_def}
\end{align}
\end{keyeqn}
This integral weights the post-switch occupation by the reference energy at $a_*$. The covariant stress-energy density needed to assess backreaction during production is calculated separately below. For the fiducial smooth profile,
$\rho_{\rm occ}^{\rm ind}/H^4\simeq5.2\times10^5$ and
$1.2\times10^8$ for $A/H=4$ and $4.5$, respectively. For both values of $A/H$, the integrand peaks near $\kappa\simeq0.037$. Increasing the tanh 10--90 width from $0.1$ to $0.8$ at fixed $\ma I_{\rm tach}(0)$ changes $\rho_{\rm occ}^{\rm ind}$ by $8.4\%$ and $2.6\%$; at width $0.4$, tanh and raised-cosine profiles with the same $\ma I_{\rm tach}(0)$ differ by $1.6\%$ and $3.1\%$. The oscillatory fine structure is more sensitive to the switching profile. We therefore compare profiles through their integrated production.

\subsection{Backreaction constraints}\label{eq:ConstraintsontheParameters}

The backreaction bounds must be satisfied throughout the production interval. Varying the covariant action gives, for a homogeneous inflaton,
\begin{align}
\rho_{\chi+{\rm int}}=
|\dot\chi|^2+\frac{|\nabla\chi|^2}{a^2}+m^2|\chi|^2
+\frac{A^2w}{2}(\chi^2+\chi^{*2})+\mu\ma J^0.
\label{eq:covariantenergyoriginal}
\end{align}
We subtract the $A=0$ reference mode by mode, imposing the same physical momentum cutoff throughout the smooth evolution. The resulting difference is a regulated, state-dependent stress. Its maximum values during amplification change negligibly when the cutoff is varied from $40H$ to $120H$.

For $w=w[\sigma]$, the derivative coupling corrects the homogeneous inflaton equation as
\begin{align}
\ddot\phi+3H\dot\phi+V_{,\phi}
=-\frac{\Delta(\nabla_\mu\ma J^\mu)}{\Lambda}.
\label{eq:inflatonbackreaction}
\end{align}
We characterize the backreaction by three quantities,
\begin{align}
f_{\rho,\max}
&=\max_N\frac{|\Delta\rho_{\chi+{\rm int}}|}
{3\mpl^2H^2},
\label{eq:defofenergyfraction}\\
r_{\phi,\max}
&=\max_N\frac{|\Delta(\nabla_\mu\ma J^\mu)|}
{3H|\dot\phi|\Lambda},
\qquad
|\delta Z_\phi|_{\max}
=\max_N\frac{|\Delta\langle\mb X^T\mb X\rangle|}{\Lambda^2}.
\label{eq:backreactiontests}
\end{align}
The last quantity is the induced correction to the inflaton kinetic term in the rotated action. Table~\ref{table:BackreactionBenchmark} gives the resulting maxima.

\begin{table}[H]
\centering
\small
\caption{Backreaction for the fiducial smooth profile. The numerical columns use $H\simeq4.5\times10^{13}\,\mathrm{GeV}$ and $\epsilon=0.002$, for which $\Lambda/H\simeq1.7\times10^3$ and $r=16\epsilon\simeq0.032$. The $A/H=4.5$ case violates the backreaction bounds.}
\label{table:BackreactionBenchmark}
\begin{tabular}{cccccc}
\toprule
$A/H$ & $\max|\Delta\rho|/H^4$ & $f_{\rho,\max}$
& $r_{\phi,\max}$ & $|\delta Z_\phi|_{\max}$ & backreaction bounds\\
\midrule
$4.0$ & $6.38\times10^4$ & $7.17\times10^{-6}$
& $2.10\times10^{-2}$ & $1.91\times10^{-2}$ & satisfied\\
$4.5$ & $1.42\times10^7$ & $1.60\times10^{-3}$
& $4.82$ & $3.94$ & violated\\
\bottomrule
\end{tabular}
\end{table}

For $A/H=4$, both inflaton corrections remain below $0.1$ when
\begin{align}
\epsilon\gtrsim4.2\times10^{-4},
\qquad\Longleftrightarrow\qquad
\Lambda/H\gtrsim7.9\times10^2
\end{align}
at the stated $H$ and $\mu/H=2$. For $A/H=4.5$, the current correction would require $\epsilon\gtrsim9.6\times10^{-2}$, incompatible with the assumed CMB-normalized single-field slow-roll regime. This conclusion is unchanged when the width of the smooth window is varied.

\begin{figure}[!t]
\centering
\includegraphics[width=0.92\textwidth]{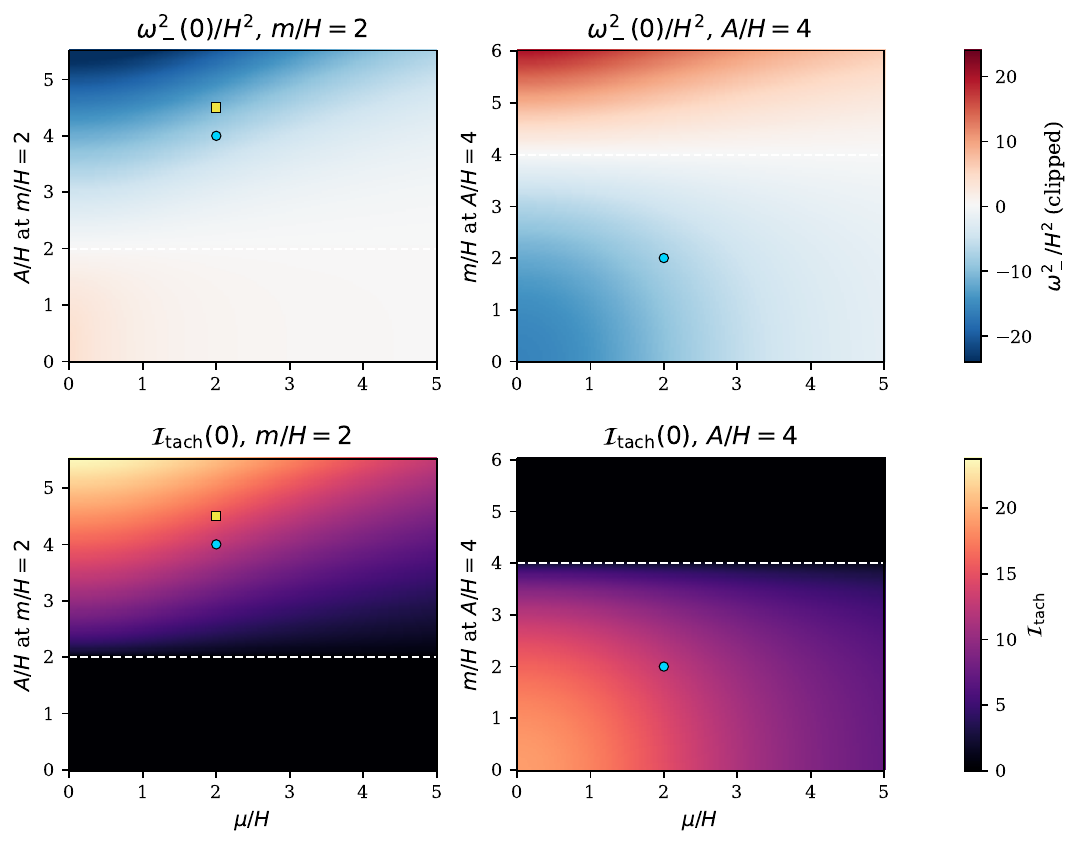}
\caption{Dependence of the local stability and tachyonic growth on the model parameters for the fiducial smooth window, $H\tau_i=-100$, $H\tau_*=-1$, and $\Delta N_{10-90}=0.2$. Top: the plateau value $\omega_-^2(q=0)/H^2$; bottom: the zero-mode growth exponent $\ma I_{\rm tach}(0)$. The left column fixes $m/H=2$ and varies $(\mu/H,A/H)$; the right fixes $A/H=4$ and varies $(\mu/H,m/H)$. White dashed lines mark $A=m$. The cyan circle denotes $(m/H,\mu/H,A/H)=(2,2,4)$, and the yellow square denotes $A/H=4.5$ in the left panels. Backreaction for the two benchmark points is given in Table~\ref{table:BackreactionBenchmark}.}
\label{fig:fx_constraints}
\end{figure}

\begin{figure}[!t]
\centering
\includegraphics[width=\textwidth]{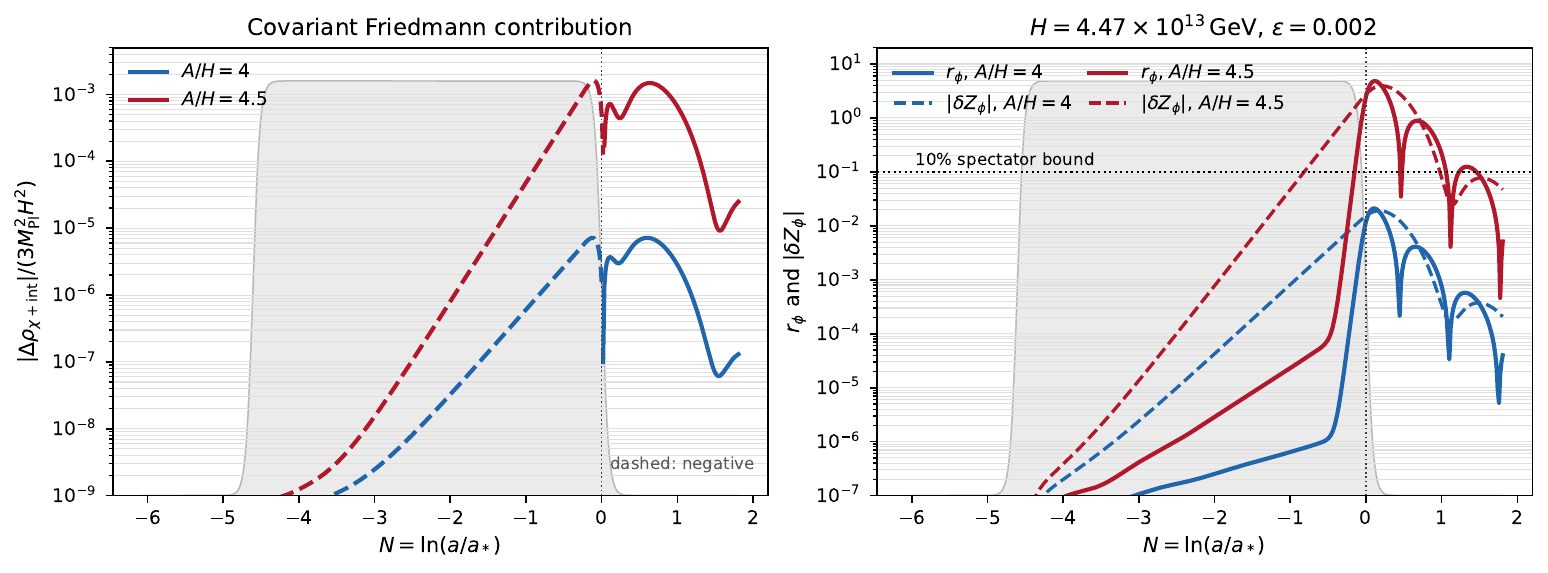}
\caption{Time dependence of the backreaction quantities for $m/H=\mu/H=2$, $H\tau_i=-100$, $H\tau_*=-1$, and $\Delta N_{10-90}=0.2$. Left: $|\Delta\rho_{\chi+{\rm int}}|/(3\mpl^2H^2)$, with dashed segments where $\Delta\rho_{\chi+{\rm int}}<0$. Right: the correction to the inflaton equation and the induced correction to its kinetic term at $H\simeq4.5\times10^{13}\,\mathrm{GeV}$ and $\epsilon=0.002$. The vertical line is the switch-off midpoint; the calculation continues through its smooth tail.}
\label{fig:fullhistorybackreaction}
\end{figure}

Figure~\ref{fig:fx_constraints} shows the plateau value of the lower branch and the smooth-window tachyonic growth exponent across the $(\mu/H,A/H)$ and $(\mu/H,m/H)$ planes. The dashed contours show the exact stability boundary $A=m$. Although the boundary is independent of $\mu$, the growth exponent varies because the chemical potential changes the magnitude of the negative eigenvalue. For the two benchmark points, the backreaction measures are given in Table~\ref{table:BackreactionBenchmark} and Fig.~\ref{fig:fullhistorybackreaction}.

The largest $|\Delta\rho_{\chi+{\rm int}}|$ occurs during switch-off; evaluating the stress tensor at the midpoint $N=0$ misses this maximum. The backreaction bound therefore requires the full time evolution of the covariant stress. The Ward identity in Appendix~\ref{app:backreactionchecks} includes the work proportional to $w_{,N}$. Determining the associated energy exchange dynamically requires a model for the trigger.

\section{Gravitational waves from particle production}\label{sec:GWPowerspectrum}

We perturb the metric according to
\begin{align}
\dd s^2=a^2(\tau)\left[-\dd\tau^2+
(\delta_{ij}+h_{ij})\dd x^i\dd x^j\right]
\end{align}
and decompose
\begin{align}
h_{ij}(\mb k,\tau)=
\sum_{\lambda=\pm}e^\lambda_{ij}(\hat{\mb k})h_\lambda(\mb k,\tau),
\qquad
e^{\lambda *}_{ij}e^{\lambda'}_{ij}=2\delta_{\lambda\lambda'}.
\end{align}
With $e^\lambda_{ii}=k_i e^\lambda_{ij}=0$, the inverse projector is
$h_\lambda=e^{\lambda *}_{ij}h_{ij}/2$. For homogeneous $\phi_0$, all non-gradient contributions to $T_{ij}^{(\chi)}$ are traces and vanish under the transverse-traceless projection. Define
\begin{align}
\pi_{ij}^{\rm cov}
&=T_{ij}^{(\chi)}-\frac13\delta_{ij}\delta^{\ell m}T_{\ell m}^{(\chi)},
\qquad
\pi_{ij}^{\rm phys}=a^{-2}\pi_{ij}^{\rm cov},
\nonumber\\
S_\lambda^{\rm phys}
&=\frac12e^{\lambda *}_{ij}\pi_{ij}^{\rm phys},
\qquad
\Sigma_\lambda\equiv a^2S_\lambda^{\rm phys}
=\frac12e^{\lambda *}_{ij}\pi_{ij}^{\rm cov}.
\label{eq:sourcedefs}
\end{align}
The tensor equation is therefore
\begin{keyeqn}
\begin{align}
h_\lambda''+2\ma Hh_\lambda'+k^2h_\lambda
=\frac{2a^2}{\mpl^2}S_\lambda^{\rm phys}
=\frac{2}{\mpl^2}\Sigma_\lambda,
\qquad \ma H\equiv\frac{a'}a.
\label{eq:EOM_Tensor_1}
\end{align}
\end{keyeqn}
With $\chi(\mb x)=\intp e^{\ii\mb p\cdot\mb x}\chi_{\mb p}$,
\begin{keyeqn}
\begin{align}
\Sigma_\lambda(\mb k,\tau)
=\intp K_\lambda(\mb p,\mb k)
\chi_{\mb p}(\tau)\chi^*_{\mb k-\mb p}(\tau),
\qquad
K_\lambda=e^{\lambda *}_{ij}p_ip_j.
\label{eq:helicalstress}
\end{align}
\end{keyeqn}
Equivalently, for $\chi=(\varphi_1+\ii\varphi_2)/\sqrt2$,
$\Sigma_\lambda=\frac12\sum_{a=1}^2\intp
K_\lambda\varphi_{a,\mb p}\varphi_{a,\mb k-\mb p}$.

The retarded Green function obeys
\begin{align}
\left(\partial_\tau^2+2\ma H\partial_\tau+k^2\right)
G_k(\tau,\tau')=\delta(\tau-\tau'),
\qquad
G_k(\tau<\tau',\tau')=0,
\label{eq:DefofGreenfunctiondS}
\end{align}
and in exact de Sitter is
\begin{align}
G_k(\tau,\tau')=
-\Theta(\tau-\tau')
\frac{(k\tau-k\tau')\cos[k(\tau-\tau')]
-(1+k^2\tau\tau')\sin[k(\tau-\tau')]}
{k^3\tau'^2}.
\label{eq:GreenfunctiondS}
\end{align}
The sourced solution is
\begin{align}
h_\lambda(\mb k,\tau)
=\frac{2}{\mpl^2}\int^\tau\dd\tau'\,
G_k(\tau,\tau')\Sigma_\lambda(\mb k,\tau').
\end{align}

\subsection{Tensor power spectrum}

Define the matrix of Wightman functions for the two real components by
\begin{align}
\langle\varphi_{a,\mb p}(\tau)\varphi_{b,\mb p'}(\tau')\rangle
=(2\pi)^3\delta^{(3)}(\mb p+\mb p')C_{ab}(p;\tau,\tau').
\end{align}
Wick's theorem gives the exact connected source kernel
\begin{keyeqn}
\begin{align}
&\langle\Sigma_\lambda(\mb k,\tau)
\Sigma_{\lambda'}(\mb k',\tau')\rangle_c=(2\pi)^3\delta^{(3)}(\mb k+\mb k')
\delta_{\lambda\lambda'}\,\ma N_\lambda[C](k;\tau,\tau'),
\nonumber\\
\ma N_\lambda[C]
&=\frac12\intp|K_\lambda|^2
\tr\!\left[C_p(\tau,\tau')C_q^T(\tau,\tau')\right],
\qquad q=|\mb k-\mb p|.
\label{eq:correlatorofthehelicalsource}
\end{align}
\end{keyeqn}
The two covariances are evaluated with the same ordering of $\tau$ and $\tau'$. In complex notation this integrand is
\begin{align}
\ma N_\lambda[C]=\intp|K_\lambda|^2
\left(F_p\bar F_q+M_p\bar M_q\right),
\label{eq:TheRenormlizedTwoPointFunc}
\end{align}
where $F=\langle\chi\chi^*\rangle'$, $\bar F=\langle\chi^*\chi\rangle'$,
$M=\langle\chi\chi\rangle'$, and $\bar M=\langle\chi^*\chi^*\rangle'$ with this same time ordering. The $M\bar M$ term is the anomalous contraction required by quadratic $U(1)$ breaking.

We calculate the excess of the connected source kernel over its $A=0$ value,
\begin{align}
\ma N_\lambda^{\rm exc}
=\ma N_\lambda[C_A]-\ma N_\lambda[C_0],
\label{eq:excesskernel}
\end{align}
where both terms are evaluated with the same regulator. For comparison, we also define the particle-particle contribution
\begin{align}
\ma N_\lambda^{\rm pp}
=\frac12\intp|K_\lambda|^2
\tr(\Delta C_p\Delta C_q^T),
\qquad \Delta C=C_A-C_0.
\label{eq:ppkernel}
\end{align}
The full excess contains two vacuum--excitation cross terms in addition to $\ma N_\lambda^{\rm pp}$. Our tensor spectra use this full excess, Eq.~\eqref{eq:excesskernel}, unless stated otherwise. An absolute renormalized stress-tensor noise kernel would additionally require the local gravitational counterterms discussed in Refs.~\cite{Parker:1974qw,Birrell:1982ix,Anderson:1987yt}.

With
\begin{align}
P_h(k,\tau)=\frac{k^3}{\pi^2}
\sum_{\lambda=\pm}\langle|h_\lambda(\mb k,\tau)|^2\rangle',
\label{eq:Phktau}
\end{align}
the tensor power obtained from this finite-time subtracted correlator is
\begin{keyeqn}
\begin{align}
P_h^{\rm exc}(k,\tau)
=\frac{2k^3}{\pi^2\mpl^4}
\sum_{\lambda=\pm}
\int^\tau\dd\tau'\int^\tau\dd\tau''\,
G_k(\tau,\tau')G_k(\tau,\tau'')
\nonumber\\
\times\intp|K_\lambda|^2
\tr\!\big[
C_{A,p}(\tau',\tau'')C_{A,q}^T(\tau',\tau'')
-C_{0,p}(\tau',\tau'')C_{0,q}^T(\tau',\tau'')
\big].
\label{eq:Phktau2}
\end{align}
\end{keyeqn}
Here $C$ is the covariance of the physical field $\chi$. A calculation using the canonical field $a\chi$ gives the same expression after dividing each two-point function by $a(\tau')a(\tau'')$.

\begin{figure}[H]
\centering
\includegraphics[width=0.62\textwidth]{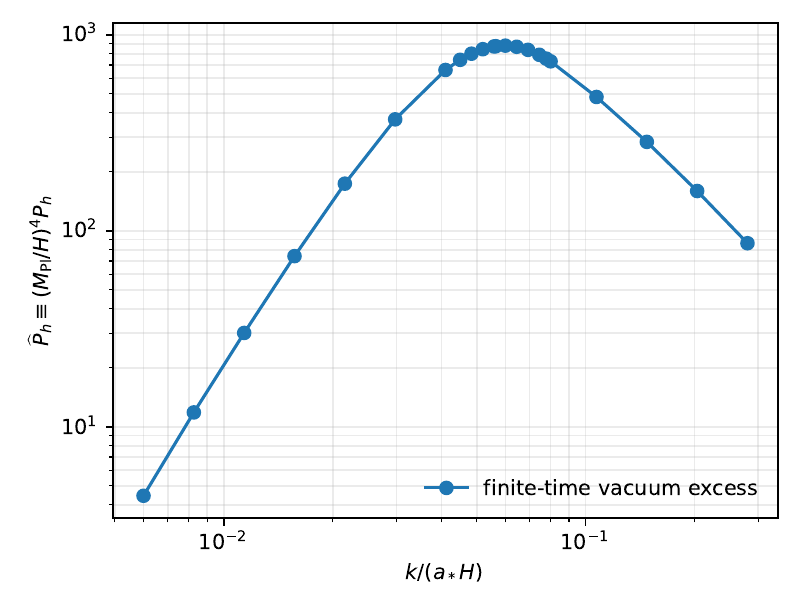}
\caption{Finite-time tensor spectrum for the $A/H=4$ benchmark with $m/H=\mu/H=2$, $H\tau_i=-100$, $H\tau_*=-1$, and $\Delta N_{10-90}=0.2$. The rescaled subtracted tensor power $(\mpl/H)^4P_h^{\rm exc}$ is obtained from the full covariance matrix of the two real fields, including normal and anomalous contractions, with identical momentum cutoffs in the interacting and $A=0$ terms.}
\label{fig:GW_Shape}
\end{figure}

We evaluate Eq.~\eqref{eq:Phktau2} using the full covariance matrix through the smooth production interval and its tail. Appendix~\ref{app:finitetime} gives the factorized expression used for the numerical integration and its convergence checks. Figure~\ref{fig:GW_Shape} shows the resulting spectrum for the $A/H=4$ benchmark.

The resulting spectrum has
\begin{align}
\frac{k_{\rm peak}}{a_*H}\simeq0.06,
\qquad
\left(\frac{\mpl}{H}\right)^4P_{h,\rm peak}^{\rm exc}\simeq8.8\times10^2.
\label{eq:correctedgwpeak}
\end{align}
At $k/(a_*H)\simeq0.06$, the particle-particle contribution agrees fractionally with the subtracted result to within $6\times10^{-5}$. This numerical agreement reflects the high occupation of the infrared modes.

For the fiducial CMB-normalized value
$H\simeq4.5\times10^{13}\,\mathrm{GeV}$ with $\epsilon=0.002$,
\begin{align}
P_{h,\rm peak}^{\rm exc}\simeq1.0\times10^{-16},
\qquad
P_h^{\rm vac}=\frac{2H^2}{\pi^2\mpl^2}
\simeq6.8\times10^{-11},
\qquad
\frac{P_{h,\rm peak}^{\rm exc}}{P_h^{\rm vac}}
\simeq1.5\times10^{-6}.
\label{eq:sourcedvacuumcomparison}
\end{align}
The sourced correction is therefore far below the standard vacuum contribution. Increasing the upper internal-momentum cutoff from $0.8a_*H$ to $a_*H$ changes the plotted spectrum by less than one part in $10^3$.

\subsection{Transfer to the present-day spectrum}

After the source has switched off, the usual tensor transfer applies. Deep inside the horizon,
\begin{align}
\Omega_{\rm GW}(k,\tau)
=\frac1{12}\left(\frac{k}{aH}\right)^2
\overline{P_h(k,\tau)}.
\label{eq:VI1}
\end{align}
For modes that re-enter during radiation domination, oscillation averaging gives
\begin{align}
\Omega_{\rm GW}(k,\tau)=\frac1{24}P_h^{\rm prim}(k).
\label{eq:VI2}
\end{align}
Including the changes in relativistic degrees of freedom,
\begin{align}
\Omega_{\rm GW,0}(k)=
\frac{\Omega_{r,0}}{24}P_h^{\rm prim}(k)
\frac{g_*(T_k)}{g_*(T_0)}
\left(\frac{g_{*s}(T_0)}{g_{*s}(T_k)}\right)^{4/3}
\ma T_\nu^2(k)\ma T_{\rm eq}^2(k).
\label{eq:VI3}
\end{align}
Equations~\eqref{eq:VI1}--\eqref{eq:VI3} describe the standard transfer of a physical primordial spectrum~\cite{Allen:1997ad,Weinberg:2003ur,Maggiore:2007ulw,Caprini:2018mtu}. The $A/H=4$ case serves as a numerical check of the finite-time calculation; its signal is too small to motivate a detector forecast [Eq.~\eqref{eq:sourcedvacuumcomparison}]. Section~\ref{sec:lisa_benchmark} considers a shorter production interval that yields a larger signal.

\subsection{Present-day frequency and relic abundance}

Let $N_{\rm post}=\ln(a_{\rm end}/a_*)$ and
$N_{\rm reh}=\ln(a_{\rm reh}/a_{\rm end})$, and define the pivot separation
$N_{\rm CMB}=\ln(a_{\rm end}/a_{\rm piv})$ for
$k_{\rm piv}=0.05\,\mathrm{Mpc}^{-1}$. For
$\kappa=k/(a_*H_*)$, entropy conservation after thermalization gives
\begin{align}
f_0(\kappa)=\frac{\kappa H_*}{2\pi}
e^{-N_{\rm post}-N_{\rm reh}}
\frac{T_0}{T_{\rm reh}}
\left(\frac{g_{*s,0}}{g_{*s,\rm reh}}\right)^{1/3}.
\label{eq:frequencydirect}
\end{align}
The equivalent relation in terms of the CMB pivot is
\begin{align}
f_0(\kappa)=f_{\rm piv}\,\kappa
\frac{H_*}{H_{\rm piv}}
\exp(N_{\rm CMB}-N_{\rm post}),
\qquad
f_{\rm piv}\simeq7.73\times10^{-17}\,\mathrm{Hz}.
\label{eq:frequencycmb}
\end{align}
A five-e-fold change in $N_{\rm CMB}$ at fixed production time therefore changes the present-day frequency by $e^5\simeq148$.

For a constant mean reheating equation of state $\bar w_{\rm reh}$,
\begin{align}
\rho_{\rm reh}=\rho_{\rm end}
e^{-3(1+\bar w_{\rm reh})N_{\rm reh}},
\qquad
T_{\rm reh}=\left(\frac{30\rho_{\rm reh}}
{\pi^2g_{*,\rm reh}}\right)^{1/4}.
\label{eq:reheatingclosure}
\end{align}
For stable $\chi$ particles that are already nonrelativistic at $a_*$ and undergo no subsequent production or annihilation, a physical energy fraction $f_{\chi,*}=\rho_{\chi,*}/\rho_*$ gives the present abundance
\begin{keyeqn}
\begin{align}
\Omega_{\chi,0}h^2
\simeq&\;2.06\times10^8
\left(\frac{g_*}{g_{*s}}\right)_{\rm reh}
\left(\frac{T_{\rm reh}}{\mathrm{GeV}}\right)
f_{\chi,*}\frac{\rho_*}{\rho_{\rm end}}
\nonumber\\
&\times
\exp[-3N_{\rm post}+3\bar w_{\rm reh}N_{\rm reh}].
\label{eq:DMrelicAbundanceMapping}
\end{align}
\end{keyeqn}
Equations~\eqref{eq:frequencydirect}--\eqref{eq:DMrelicAbundanceMapping} relate the production time and abundance to CMB matching and the reheating history. These quantities must be specified consistently.

\begin{table}[H]
\centering
\small
\caption{Present-day frequencies of the $A/H=4$ smooth profile evaluated at three values of $N_{\rm CMB}$, at fixed $N_{\rm post}\simeq16.6$ and $H_*=H_{\rm piv}$. No abundance is assumed. The final column uses the tensor peak.}
\label{table:Benchmark}
\begin{tabular}{ccc}
\toprule
$N_{\rm CMB}$ & $f_0(\kappa=1)$ [Hz] & $f_0(\kappa_{\rm peak})$ [Hz]\\
\midrule
$50$ & $2.47\times10^{-2}$ & $1.48\times10^{-3}$\\
$55$ & $3.67$ & $0.219$\\
$60$ & $5.45\times10^2$ & $32.5$\\
\bottomrule
\end{tabular}
\end{table}

The occupation-weighted integral in Eq.~\eqref{eq:rhop_def} is not a renormalized physical density and has not been evolved through reheating. Table~\ref{table:Benchmark} therefore illustrates only the frequency dependence of the $A/H=4$ profile. An abundance calculation would additionally depend on any period during which $\chi$ remains relativistic, entropy transfer, interactions, instability, and the fraction of dark matter carried by $\chi$.

\subsection{LISA-band benchmarks}\label{sec:lisa_benchmark}

Using the same action with a shorter tanh production window, we find two nearby parameter points with much larger tensor signals than the $A/H=4$ case:
\begin{align}
\frac{A}{H}\simeq10.5,\qquad \frac{m}{H}=2,
\qquad N_*-N_i=1,\qquad \Delta N_{10-90}=0.6,
\qquad \frac{\mu}{H}=0.036\ \hbox{or}\ 0.038.
\end{align}
Both points lie in the finite-duration tachyonic regime, $A>m$. We choose
\begin{align}
r=8\left(\frac{\mu}{H}\right)^2
\simeq(0.0104,\,0.0116),
\qquad \epsilon=\frac r{16},
\qquad \Lambda=\mpl,
\end{align}
where $\mu/H=\sqrt{r/8}\,\mpl/\Lambda$. With $\ma P_\zeta=2.1\times10^{-9}$ this gives $H\simeq(2.5,\,2.7)\times10^{13}\,\mathrm{GeV}$. The two values of $r$ are below the combined Planck--BICEP/Keck bound $r<0.032$~\cite{Tristram:2021tvh}.

For $\mu/H=0.038$, which gives the larger backreaction, the maxima of the canonical and covariant Friedmann fractions, the inflaton-current force ratio, the correction to the inflaton kinetic term, and the normalized trigger work per Hubble time are, respectively,
\begin{align}
(1.2,\ 1.2,\ 1.8,\ 1.8,\ 8.0)\%.
\end{align}
The largest ratio at $\mu/H=0.036$ is $7.2\%$, so both benchmark points satisfy the $10\%$ criterion. At $\mu/H=0.038$, the rms field amplitude reaches $\sqrt{\langle\mb X^T\mb X\rangle}/\mpl\simeq0.14$. The ratios $A/\Lambda\simeq1.2\times10^{-4}$, $m/\Lambda\simeq2.2\times10^{-5}$, and $\mu/\Lambda\simeq4.2\times10^{-7}$ place the physical scales well below the derivative-coupling cutoff. These results are consistent with a perturbative spectator treatment of $\chi$ for the prescribed profile; the trigger dynamics remains to be studied.

For $\mu/H=0.038$, the particle-particle result in Eq.~\eqref{eq:ppkernel} peaks at
\begin{align}
\frac{k_{\rm peak}}{a_*H}\simeq1.7,
\qquad P_{h,\rm peak}\simeq1.8\times10^{-6}.
\end{align}
Over the calculated range of wavenumbers, $P_h^{\rm pp}/P_h^{\rm exc}$ differs from unity by less than $9.9\times10^{-7}$, so the particle-particle and subtracted spectra agree to this accuracy. This comparison uses regulated, state-dependent correlators; absolute renormalization of the stress-tensor noise kernel lies beyond the present calculation. Applying Eq.~\eqref{eq:VI3} for high-temperature radiation re-entry, with $\ma T_\nu^2=0.645$, gives Table~\ref{table:DetectableBenchmark}.

\begin{table}[H]
\centering
\small
\caption{Results for the two LISA-band benchmarks. The conversion to present-day frequency assumes instantaneous reheating and $H_*=H_{\rm end}$; the final row includes the conservative compact-binary foreground treatment described in the text.}
\label{table:DetectableBenchmark}
\begin{tabular}{lcc}
\toprule
Quantity & $\mu/H=0.036$ & $\mu/H=0.038$\\
\midrule
$r$ & $0.0104$ & $0.0116$\\
$H$ [GeV] & $2.52\times10^{13}$ & $2.66\times10^{13}$\\
maximum backreaction ratio & $0.0717$ & $0.0799$\\
$P_{h,\rm peak}$ & $1.43\times10^{-6}$ & $1.77\times10^{-6}$\\
$\Omega_{\rm GW,0}^{\rm peak}$ & $1.36\times10^{-12}$ & $1.69\times10^{-12}$\\
$h^2\Omega_{\rm GW,0}^{\rm peak}$ & $6.20\times10^{-13}$ & $7.70\times10^{-13}$\\
$N_{\rm CMB}$ & $55.64$ & $55.66$\\
$N_{\rm post}$ & $24.51$ & $24.54$\\
$f_{\rm peak}$ [mHz] & $4.23$ & $4.23$\\
four-year SNR including the compact-binary foreground & $10.1$ & $12.5$\\
\bottomrule
\end{tabular}
\end{table}

Table~\ref{table:DetectableBenchmark} assumes instantaneous reheating with $g_*=g_{*s}=106.75$ and $H_*=H_{\rm end}$. For this post-inflationary history, CMB matching gives the quoted $N_{\rm CMB}$ and $N_{\rm post}$, and Eqs.~\eqref{eq:frequencydirect} and~\eqref{eq:frequencycmb} give the same numerical frequencies. The calculated spectrum spans approximately $1.3$--$13\,\mathrm{mHz}$. A different reheating history would shift the required production time and hence the observed frequency.

\begin{figure}[H]
\centering
\includegraphics[width=0.98\textwidth]{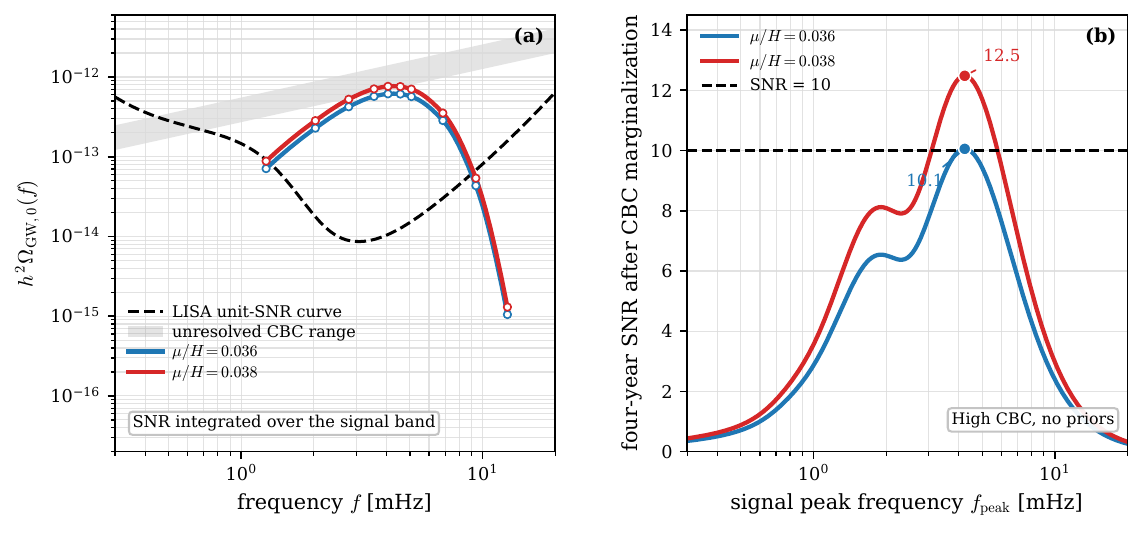}
\caption{Present-day spectra and LISA forecast for the two $\Lambda=\mpl$ benchmark points, with $A/H\simeq10.5$, $m/H=2$, $N_*-N_i=1$, and $\Delta N_{10-90}=0.6$. (a) Neutrino-damped particle-particle spectra mapped to the frequencies that maximize the SNR after foreground marginalization. Points show the numerical results, and the curves interpolate between them. The dashed curve is the combined instrumental and Galactic-confusion noise curve corresponding to unit SNR per logarithmic frequency interval, and the gray band shows the interquartile range of the unresolved compact-binary amplitude. (b) Four-year SNR after including a compact-binary background at the upper edge of this interquartile range and marginalizing over its amplitude and spectral slope. The two cases give SNR $10.1$ and $12.5$ at $f_{\rm peak}\simeq4.2\,\mathrm{mHz}$.}
\label{fig:DetectableGW}
\end{figure}

For an idealized LISA forecast, we use the standard $2.5\,\mathrm{Gm}$, six-link analytic noise model and its four-year fit to the Galactic confusion noise~\cite{Robson:2018ifk}, integrating only over the calculated signal band. For a known spectral shape, the signal-to-noise ratio is
\begin{align}
{\rm SNR}^2=T_{\rm obs}\int\dd f\,
\left[\frac{\Omega_{\rm GW,0}(f)}{\Omega_n(f)}\right]^2,
\qquad T_{\rm obs}=4\ {\rm yr}.
\end{align}
We vary the production time to maximize the final statistic after foreground marginalization. At the resulting peak frequencies, the known-shape SNR from instrumental and Galactic confusion noise is about $52$ and $64$. Projecting out the compact-binary amplitude and spectral slope without priors lowers these values to about $11$ and $14$. Finally, we include a compact-binary background at the upper end of its interquartile range, $h^2\Omega_{\rm CBC}(3\,\mathrm{mHz})=1.15\times10^{-12}$~\cite{Babak:2023lro}, in the noise covariance and again marginalize over the same two parameters. This conservative choice gives SNR $10.1$ and $12.5$ at $f_{\rm peak}\simeq4.2\,\mathrm{mHz}$, as shown in Fig.~\ref{fig:DetectableGW}.

Under these assumptions, the $\mu/H=0.038$ benchmark lies within the reach of this idealized LISA forecast. These points come from a limited parameter scan. At leading order in the decoupling limit, the direct contribution to the scalar power is $\ma P_{\zeta,\rm direct}^{\rm peak}\simeq(1.5,\,1.8)\times10^{-4}$ for the two points, corresponding to rms curvature perturbations of order $10^{-2}$. Assessing the non-Gaussian tail, isocurvature, and trigger fluctuations requires a complete gauge-invariant scalar calculation beyond this two-point estimate. A robust detection forecast would also require a full LISA foreground likelihood and an absolute renormalization of the stress-tensor noise kernel.

\section{Relation to cosmological-collider calculations}\label{sec:SecofCC}

The derivative coupling and quadratic breaking operator also arise in cosmological-collider studies~\cite{Bodas:2020yho,ArkaniHamed2015,Chen2010,Hubisz:2024xnj,Lee2016,Chen2017,Cui2022,Wang2020,Wang2020a,Chen2018,Chen:2016hrz,Chen:2016uwp,Chen:2016nrs,Chen:2015lza,Tong:2022cdz,Pimentel2022,Yin:2023jlv,Aoki:2020zbj,Pinol:2021aun,Aoki:2023tjm,Aoki:2025uff,Aoki:2024jha}. We compare the charge normalization and parameter regimes of these calculations: Ref.~\cite{Bodas:2020yho} treats perturbative small breaking in an extended effective theory, while the tensor benchmarks in Sections~\ref{sec:particleproduction} and~\ref{sec:GWPowerspectrum} use the minimal action in Eq.~\eqref{eq:lagcorvariant} at large breaking.

The phase of the breaking operator has charge two. Expanding it around the inflaton background gives
\begin{keyeqn}
\begin{align}
e^{-2\ii(\phi_0+\delta\phi)/\Lambda}
=e^{-2\ii\phi_0/\Lambda}
\left[
1-\frac{2\ii\,\delta\phi}{\Lambda}
-\frac{2\,\delta\phi^2}{\Lambda^2}
+\cdots
\right].
\label{eq:ccphase}
\end{align}
\end{keyeqn}
The background phase therefore rotates at $2\mu$, the one-$\delta\phi$ vertex carries $-2\ii/\Lambda$, and the quadratic vertex carries $-2/\Lambda^2$. This charge-two normalization agrees with the convention in Ref.~\cite{Bodas:2020yho}.

\begin{figure}[H]
\centering
\includegraphics[width=0.78\textwidth]{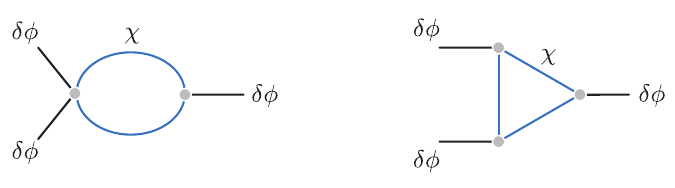}
\caption{Loop topologies contributing to the bispectrum in the perturbative small-$U(1)$-breaking calculation of Ref.~\cite{Bodas:2020yho}.}
\label{fig:FeynmannDiagram}
\end{figure}

The conventional dimensionless bispectrum shape is
\begin{align}
S(k_1,k_2,k_3)=\frac56
\frac{\langle\zeta_{\mb k_1}\zeta_{\mb k_2}\zeta_{\mb k_3}\rangle'}
{P_\zeta(k_1)P_\zeta(k_2)+P_\zeta(k_2)P_\zeta(k_3)
+P_\zeta(k_3)P_\zeta(k_1)}.
\label{eq:generalshapedef}
\end{align}
The squeezed-limit calculation treats the quadratic breaking through
perturbative insertions. In the notation of Ref.~\cite{Bodas:2020yho}, an
additional symmetry-preserving operator with coefficient $c$ gives
$M_{\rm eff}^2=M^2+(1-c)\lambda^2$. Its narrow-resonance hierarchy is
\begin{align}
\lambda>M_{\rm eff}\gg m_B>H,
\qquad
\varepsilon_{\rm ins}\sim
\frac{m_B^4}{H\lambda|\lambda^2-M_{\rm eff}^2|}\ll1,
\label{eq:ccinsertionscope}
\end{align}
where $m_B$ is the breaking mass and $\lambda$ is the chemical scale. In the minimal action, $m_B\to A$, $\lambda\to\mu$, and $c=0$, so
\begin{align}
M_{\rm eff}^{(0)}=\sqrt{m^2+\mu^2}>\mu
\end{align}
for positive $m^2$; the required condition $\lambda>M_{\rm eff}$ cannot be met. The insertion parameter $A^4/[H\mu|\mu^2-(M_{\rm eff}^{(0)})^2|]$ is approximately $32$ and $51$ for $A/H=4$ and $4.5$, respectively, at $m/H=\mu/H=2$, well outside the perturbative regime. Both cases have a tachyonic branch, and the latter also violates the homogeneous backreaction bounds. The $A/H\simeq10.5$ benchmark lies still farther outside the perturbative expansion in insertions of the breaking operator. The small-breaking collider formula therefore cannot be extrapolated to these benchmarks; a dedicated large-$A$ in-in calculation would be required.

\section{Conclusions and outlook}\label{sec:conclusions}

We have studied heavy-particle production during inflation in the presence of a quadratic $U(1)$-breaking operator and the derivative-current scale $\mu=\dot\phi_0/\Lambda$. The basis-independent dispersion relation develops a tachyonic branch for $q^2<A^2-m^2$. The chemical potential changes the growth exponent, the rotating eigenvectors, and the charge asymmetry, while the stability boundary remains $A=m$. Evolutions with smooth switching, comparisons between profiles with the same growth exponent, and calculations with the background held fixed identify tachyonic growth as the source of the amplification for $A/H=4$ and $4.5$.

The $A/H=4$ and $4.5$ benchmarks have different backreaction constraints. At $H\simeq4.5\times10^{13}\,\mathrm{GeV}$ and $\epsilon=0.002$, the $A/H=4$ case gives $f_{\rho,\max}\simeq7.2\times10^{-6}$, $r_{\phi,\max}\simeq0.021$, and $|\delta Z_\phi|_{\max}\simeq0.019$; it satisfies the bounds for $\epsilon\gtrsim4.2\times10^{-4}$. The $A/H=4.5$ case instead develops order-unity inflaton-current and kinetic corrections despite its small Friedmann fraction. The backreaction bounds must therefore be checked throughout the switching interval.

For $A/H=4$, the full covariance matrix of the two real fields gives a subtracted tensor spectrum peaking at $k/(a_*H)\simeq0.06$. At the fiducial CMB normalization, the sourced peak is only $1.5\times10^{-6}$ of the standard vacuum tensor power.

Within the same action, we find two $\Lambda=\mpl$ benchmarks for which all five homogeneous backreaction measures remain below $10\%$. The $\mu/H=0.038$ point gives $\Omega_{\rm GW,0}^{\rm peak}\simeq1.7\times10^{-12}$. Assuming instantaneous reheating, its peak lies at $4.2\,\mathrm{mHz}$ and gives a four-year LISA SNR of $12.5$ after marginalizing over the compact-binary foreground amplitude and spectral slope; the neighboring $\mu/H=0.036$ point gives SNR $10.1$. For the prescribed switching profile, the larger of these signals could be accessible to LISA\@. Its predicted frequency depends on the production time, CMB matching, and reheating history. The quoted SNR uses the two-parameter foreground model described in Section~\ref{sec:lisa_benchmark}.

The small-breaking cosmological-collider result reviewed in Section~\ref{sec:SecofCC} belongs to an extended EFT with $\lambda>M_{\rm eff}$. The minimal $c=0$ action has $M_{\rm eff}^{(0)}>\mu$. Predicting a cosmological-collider signal for the large-$A$ tensor benchmarks therefore requires a separate calculation.

Several questions remain to be addressed. A dynamical trigger sector should replace the prescribed profile and track the associated energy transfer. Lattice evolution can test rescattering and nonlinear backreaction at large occupation~\cite{Caravano:2022epk}; sufficiently rapid spectator scattering could also lead to a dissipative regime related to warm inflation~\cite{Berera:1995ie,Berera:2008ar,Ferreira:2017lnd,Berghaus:2019whh,Berghaus:2025dqi}. A reliable assessment of detectability will require a fully renormalized stress-tensor noise kernel, a gauge-invariant scalar calculation including the non-Gaussian tail, and a multi-component LISA likelihood.

\section*{Acknowledgments}

We thank Seung J. Lee and Chon Man Sou for useful discussions and comments on an earlier version of this manuscript. Research of Y.Y. is supported by the KIAS New Generation Research Program under Grant No.~PG105801.

\appendix

\section{Evolution of the Bogoliubov coefficients}\label{sec:Bogliubov}

\begin{figure}[H]
\centering
\includegraphics[width=\textwidth]{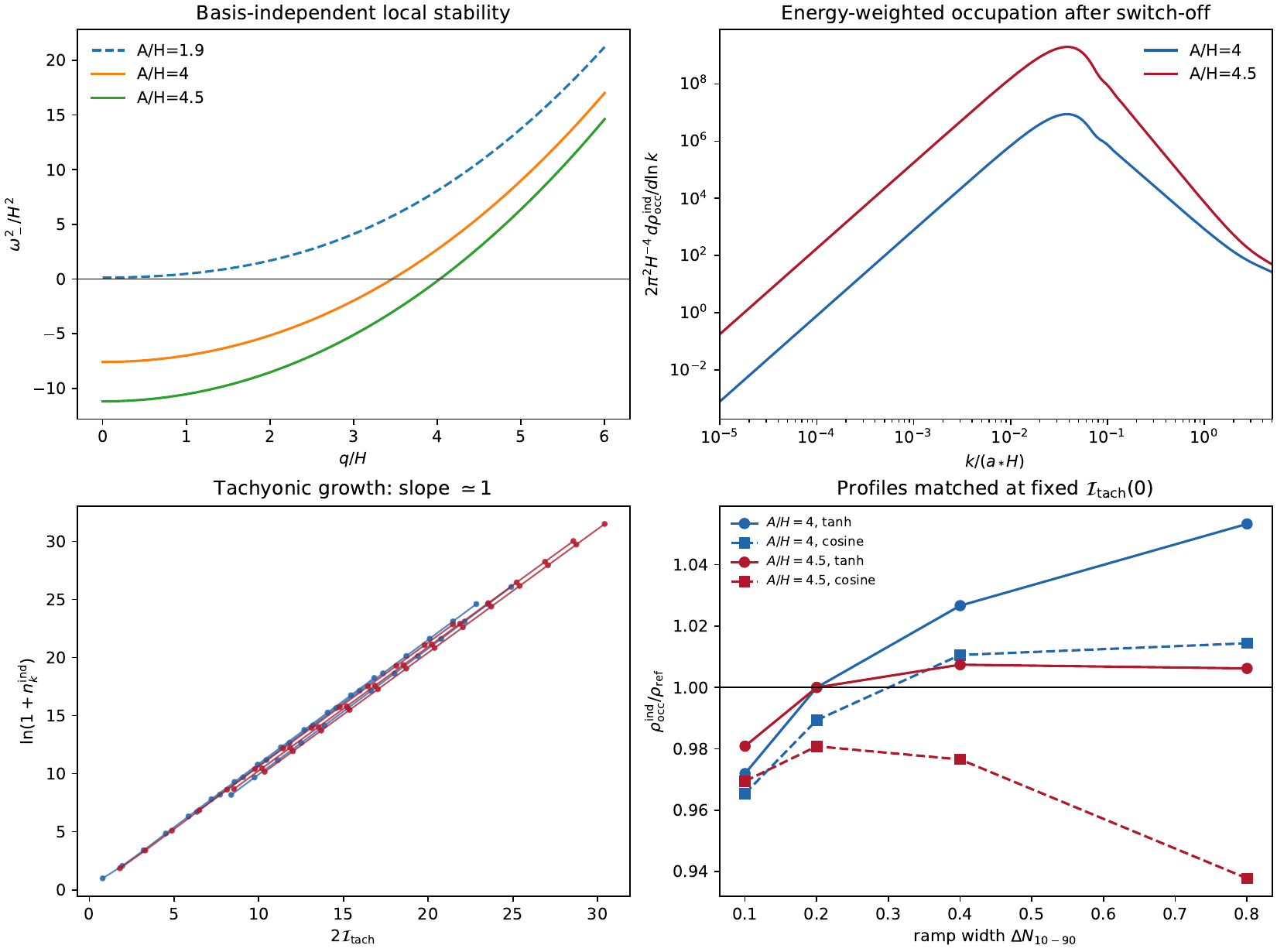}
\caption{Relation between the local instability and particle production for $m/H=\mu/H=2$, $H\tau_i=-100$, $H\tau_*=-1$, and the fiducial width $\Delta N_{10-90}=0.2$. The large-breaking curves use $A/H=4$ and $4.5$, while $A/H=1.9$ is a stable case. Top left: squared lower normal-mode frequency. Top right: energy-weighted occupation spectrum after switching. Bottom left: relation to the tachyonic growth exponent. Bottom right: tanh and raised-cosine profiles with the same zero-mode growth exponent.}
\label{fig:mechanismaudit}
\end{figure}

Let $f_k(N)$ be the exact free de Sitter reference mode for the rescaled field $\mb Y$, normalized by
\begin{align}
f_k f_{k,N}^*-f_k^*f_{k,N}=\ii.
\end{align}
Writing
\begin{align}
\varrho=\sqrt{\frac{m^2+\mu^2}{H^2}-\frac94},
\qquad z=\kappa e^{-N},
\end{align}
the Bunch--Davies reference is
\begin{align}
f_k(N)=\frac{\sqrt\pi}{2}
\exp\!\left[\frac{\ii\pi}{4}(2\ii\varrho+1)\right]
H_{\ii\varrho}^{(1)}(z).
\label{eq:exactbdmode}
\end{align}
The initial matrices are $\mb F_k=f_k\mb1$ and
$\mb V_k=f_{k,N}\mb1$. After the smooth profile has switched off, projection onto the same reference gives
\begin{align}
\mb\alpha_k
&=\ii(f_k^*\mb V_k-f_{k,N}^*\mb F_k),
\nonumber\\
\mb\beta_k
&=\ii(f_{k,N}\mb F_k-f_k\mb V_k).
\label{eq:matrixprojection}
\end{align}
For $A=0$, $\mb\beta_k$ vanishes by construction. A further Bogoliubov transformation to a future adiabatic out state may be applied separately. Figures~\ref{fig:TheNumberDensity} and~\ref{fig:mechanismaudit} use the occupation relative to the $A=0$ solution defined in Eq.~\eqref{eq:occupationmatrix}.

The fiducial DOP853 spectra satisfy the $A=0$, basis-rotation, and projection-time checks, with relative changes in $\rho_{\rm occ}^{\rm ind}$ under combined numerical variations no larger than $2.0\times10^{-6}$ and dominant-mode occupations agreeing with the 60-digit Richardson-extrapolated symplectic-midpoint result to better than $10^{-8}$.

Figure~\ref{fig:mechanismaudit} relates the large-$A$ spectra to the lower normal-mode frequency and the tachyonic growth exponent. The stable $A/H=1.9$ case and the comparison of smooth profiles test the sensitivity of particle production to the switching procedure.

\begin{figure}[H]
\centering
\includegraphics[width=\textwidth]{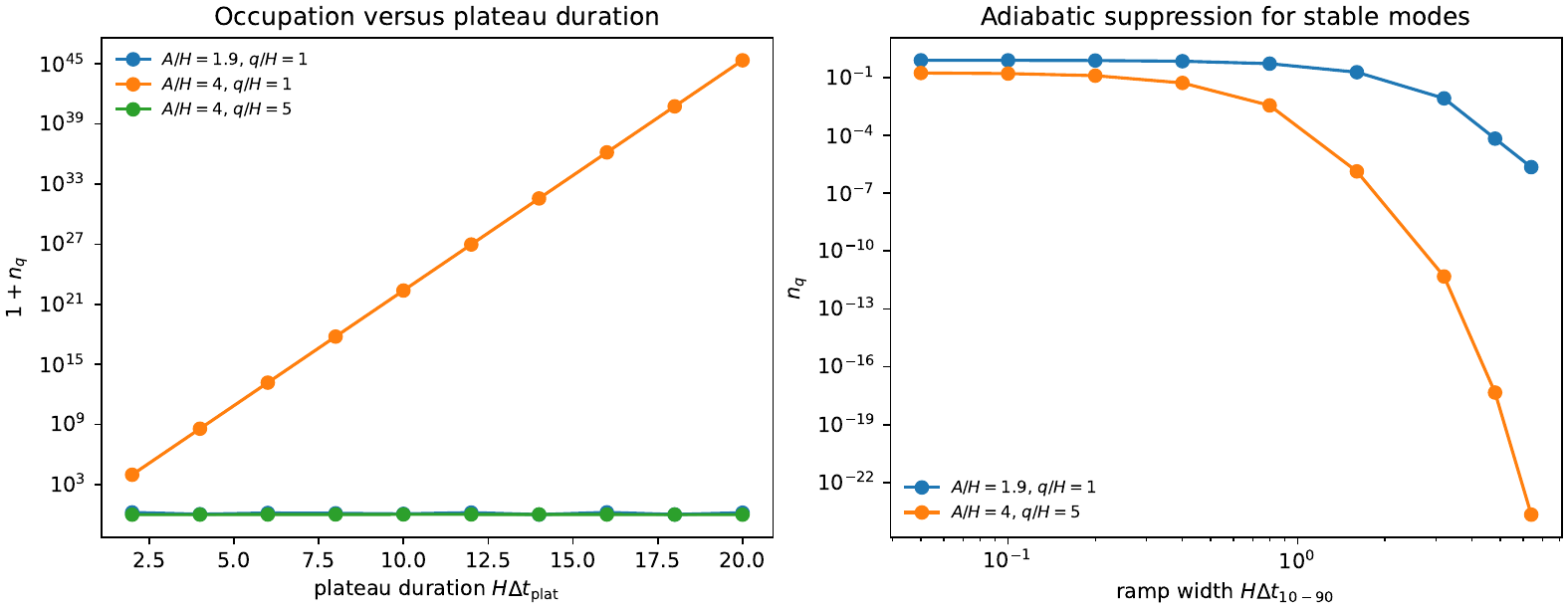}
\caption{Fixed-background calculation for $m/H=\mu/H=2$. Left: exponential growth occurs only for a mode inside the tachyonic band; for $A/H=4$ and $q/H=1$, the fitted slope agrees with $2\sqrt7$ to better than one part in $10^6$. Right: particle production in the stable mode decreases as the switching ramp becomes adiabatic.}
\label{fig:minkowskicontrols}
\end{figure}

\section{Finite-time tensor calculation and numerical convergence}\label{app:finitetime}

Set $a_*=H=1$ for the dimensionless numerical formulae and write
$\kappa=k/(a_*H)$, $\varpi=p/(a_*H)$, and
$r=\sqrt{\kappa^2+\varpi^2-2\kappa\varpi s}$, where $s=\hat{\mb p}\cdot\hat{\mb k}$. Let $\mb F_{X,\varpi}$ denote the rescaled-field mode matrix defined in Eq.~\eqref{eq:matrixmode}, evolved with the breaking interaction ($X=A$) or in the reference theory ($X=0$). At a common finite regulator, the late-time limit of Eq.~\eqref{eq:Phktau2} is evaluated as the difference of the factorized spectra
\begin{align}
\widehat P_h^X(\kappa)
&\equiv\left(\frac{\mpl}{H}\right)^4P_h^X(\kappa)
\nonumber\\
&=\frac{\kappa}{2\pi^4}
\int_{\varpi_{\rm IR}}^{\varpi_{\rm UV}}\dd\varpi\,\varpi^6
\int_{-1}^{1}\dd s\,(1-s^2)^2
\mb1_{[\varpi_{\rm IR},\varpi_{\rm UV}]}(r)
\sum_{I,J=1}^2|\ma I^X_{IJ}|^2,
\label{eq:dimensionlessexactpower}\\
\ma I^X_{IJ}
&=\int_{N_a}^{N_b}\dd N\,e^{-4N}
j_1(\kappa e^{-N})
[\mb F^T_{X,\varpi}(N)\mb F_{X,r}(N)]_{IJ},
\qquad X\in\{A,0\}.
\end{align}
Here $j_1(z)=\sin z/z^2-\cos z/z$ is the spherical Bessel function of order one. The $A/H=4$ spectrum in Fig.~\ref{fig:GW_Shape} uses $N_a=-6.2$, $N_b=1.5$, the common internal-momentum range
$0.001\leq\varpi,r\leq0.8$, and a $110\times72$ momentum--angular
quadrature. The same time interval and momentum cutoffs are imposed on the $A$ and
$A=0$ terms before their difference is formed.
The $e^{-4N}$ weight follows from the conformal-time measure and the two physical-field factors. The $I,J$ sum includes both real degrees of freedom. To avoid loss of numerical precision, we rewrite the signed difference
$\widehat P_h^{\rm exc}=\widehat P_h^A-\widehat P_h^0$ as
\begin{align}
|\ma I^A|^2-|\ma I^0|^2
=2\operatorname{Re}[(\ma I^0)^*(\ma I^A-\ma I^0)]
+|\ma I^A-\ma I^0|^2.
\end{align}

For the full covariance calculation, the $90\times64$ and $140\times90$ momentum--angular quadrature grids change the grid point nearest the peak by
$-3.50\times10^{-6}$ and $+3.91\times10^{-6}$ relative to the displayed $110\times72$ grid. Raising the infrared boundary to $0.002$ or $0.005$ changes it by at most $2.63\times10^{-6}$, and halving the time-grid density changes it by $5.50\times10^{-7}$. Using the same $130\times84$ grid, extending the common internal-momentum range from $0.8$ to $1.0$ changes the point near the peak by $-2.69\times10^{-7}$ and the point at the upper edge of the plotted $\kappa$ range by $+7.68\times10^{-4}$.

\section{Homogeneous backreaction and numerical checks}\label{app:backreactionchecks}

In the rotated real basis, with
$\mb O=\mb R\mb\sigma_3\mb R^T$,
$\mb C_\kappa=\mb F_\kappa\mb F_\kappa^\dagger$, and
$\mb D_\kappa=\partial_N\mb F_\kappa-3\mb F_\kappa/2$, the fixed physical
cutoff $q_{\rm UV}$ corresponds to
$\kappa_{\rm UV}(N)=(q_{\rm UV}/H)e^N$. All momentum integrals in this
appendix use this same time-dependent upper limit in both the $A$ and $A=0$
terms. The common-cutoff energy excess is
\begin{align}
\frac{\Delta\rho_{\chi+{\rm int}}}{H^4}
=\frac{e^{-3N}}{4\pi^2}
\int^{\kappa_{\rm UV}(N)}\dd\kappa\,\kappa^2
\Delta\bigg\{
&\tr(\mb D_\kappa\mb D_\kappa^\dagger)
\nonumber\\
&+\left[\kappa^2e^{-2N}
+\frac{m^2-\mu^2}{H^2}\right]\tr\mb C_\kappa
+\frac{A^2}{H^2}w\,\tr(\mb O\mb C_\kappa)
\bigg\}.
\label{eq:timeresolvedrho}
\end{align}
Including the derivative-interaction energy in the Friedmann source gives the coefficient $m^2-\mu^2$. The occupation measure in Eq.~\eqref{eq:rhop_def} uses the positive reference Hamiltonian with mass squared $m^2+\mu^2$.

Introducing $\mb J_2=\left(\begin{smallmatrix}0&-1\\1&0\end{smallmatrix}\right)$, the current divergence is
\begin{align}
\frac{\Delta(\nabla_\mu\ma J^\mu)}{H^4}
=\frac{e^{-3N}}{2\pi^2}\int\dd\kappa\,\kappa^2\Delta\bigg[
&-2\frac{\mu}{H}\operatorname{Re}
\tr(\mb F_\kappa\mb D_\kappa^\dagger)
-3\frac{\mu}{H}\tr\mb C_\kappa
\nonumber\\
&+\frac{A^2}{H^2}w\,
\tr(\mb J_2\mb O\mb C_\kappa)
\bigg].
\label{eq:modecurrentdivergence}
\end{align}
The remaining bilinear is
\begin{align}
\frac{\Delta\langle\mb X^T\mb X\rangle}{H^2}
=\frac{e^{-3N}}{2\pi^2}
\int\dd\kappa\,\kappa^2\Delta\tr\mb C_\kappa.
\end{align}
With $Q=\langle\mb X^T\mb O\mb X\rangle$, these quantities obey
\begin{align}
\Delta\rho_{,N}+3\Delta(\rho+p)
=\frac{\mu}{H}\Delta(\nabla_\mu\ma J^\mu)
+\frac{A^2}{2}w_{,N}\Delta Q.
\label{eq:energyward}
\end{align}
For the action with $w[\sigma]$, the last term is work exchanged with the trigger field. A profile $w(\phi)$ would introduce an additional force in the inflaton equation,
$-(A^2/2)w_{,\phi}\Delta Q$.

For the numerical checks we use 241 logarithmic momenta, 481 time samples, a common physical cutoff, and solver tolerances
$(4\times10^{-10},4\times10^{-12})$. The maximum difference between the covariant energies evaluated in the original and rotated bases, divided by the larger of their magnitudes and unity, is $8\times10^{-16}$; the analytic current divergence agrees with a direct derivative to better than $9\times10^{-4}$; and the Ward-identity residual, normalized to the maximum magnitude of either side of Eq.~\eqref{eq:energyward}, is at most $1.8\times10^{-4}$. Varying the common physical cutoff from $40H$ to $120H$ changes the maxima during amplification by less than $2\times10^{-6}$ in relative terms.

\bibliography{refs}

@article{Adshead:2019aac,
    author = "Adshead, Peter and Pearce, Lauren and Peloso, Marco and Roberts, Michael A. and Sorbo, Lorenzo",
    title = "{Gravitational waves from fermion production during axion inflation}",
    eprint = "1904.10483",
    archivePrefix = "arXiv",
    primaryClass = "astro-ph.CO",
    doi = "10.1088/1475-7516/2019/10/018",
    journal = "JCAP",
    volume = "10",
    pages = "018",
    year = "2019"
}

@article{Kolb:1998ki,
    author = "Kolb, Edward W. and Chung, Daniel J. H. and Riotto, Antonio",
    editor = "Falomir, H. and Gamboa Saravi, R. E. and Schaposnik, F. A.",
    title = "{WIMPzillas!}",
    eprint = "hep-ph/9810361",
    archivePrefix = "arXiv",
    reportNumber = "FERMILAB-CONF-98-325-A",
    doi = "10.1063/1.59655",
    journal = "AIP Conf. Proc.",
    volume = "484",
    number = "1",
    pages = "91--105",
    year = "1999"
}

@article{Berera:2008ar,
    author = "Berera, Arjun and Moss, Ian G. and Ramos, Rudnei O.",
    title = "{Warm Inflation and its Microphysical Basis}",
    eprint = "0808.1855",
    archivePrefix = "arXiv",
    primaryClass = "hep-ph",
    doi = "10.1088/0034-4885/72/2/026901",
    journal = "Rept. Prog. Phys.",
    volume = "72",
    pages = "026901",
    year = "2009"
}

@article{Berghaus:2019whh,
    author = "Berghaus, Kim V. and Graham, Peter W. and Kaplan, David E.",
    title = "{Minimal Warm Inflation}",
    eprint = "1910.07525",
    archivePrefix = "arXiv",
    primaryClass = "hep-ph",
    doi = "10.1088/1475-7516/2020/03/034",
    journal = "JCAP",
    volume = "03",
    pages = "034",
    year = "2020",
    note = "[Erratum: JCAP 10, E02 (2023)]"
}

@article{Caprini:2018mtu,
    author = "Caprini, Chiara and Figueroa, Daniel G.",
    title = "{Cosmological Backgrounds of Gravitational Waves}",
    eprint = "1801.04268",
    archivePrefix = "arXiv",
    primaryClass = "astro-ph.CO",
    doi = "10.1088/1361-6382/aac608",
    journal = "Class. Quant. Grav.",
    volume = "35",
    number = "16",
    pages = "163001",
    year = "2018"
}

@article{Bodas:2024hih,
    author = "Bodas, Arushi and Broadberry, Edward and Sundrum, Raman",
    title = "{Grand unification at the cosmological collider with chemical potential}",
    eprint = "2409.07524",
    archivePrefix = "arXiv",
    primaryClass = "hep-ph",
    reportNumber = "FERMILAB-PUB-24-0568-V",
    doi = "10.1007/JHEP01(2025)115",
    journal = "JHEP",
    volume = "01",
    pages = "115",
    year = "2025"
}

@article{Bodas:2025wuk,
    author = "Bodas, Arushi and Broadberry, Edward and Sundrum, Raman and Xu, Zhaohui",
    title = "{Charged Loops at the Cosmological Collider with Chemical Potential}",
    eprint = "2507.22978",
    archivePrefix = "arXiv",
    primaryClass = "hep-ph",
    reportNumber = "FERMILAB-PUB-25-0519-V",
    month = "7",
    year = "2025"
}

@article{Allen:1997ad,
    author = "Allen, Bruce and Romano, Joseph D.",
    title = "{Detecting a stochastic background of gravitational radiation: Signal processing strategies and sensitivities}",
    eprint = "gr-qc/9710117",
    archivePrefix = "arXiv",
    reportNumber = "WISC-MILW-97-TH-14",
    doi = "10.1103/PhysRevD.59.102001",
    journal = "Phys. Rev. D",
    volume = "59",
    pages = "102001",
    year = "1999"
}

@book{Maggiore:2007ulw,
    author = "Maggiore, Michele",
    title = "{Gravitational Waves. Vol. 1: Theory and Experiments}",
    doi = "10.1093/acprof:oso/9780198570745.001.0001",
    isbn = "978-0-19-171766-6, 978-0-19-852074-0",
    publisher = "Oxford University Press",
    year = "2007"
}

@article{Weinberg:2003ur,
    author = "Weinberg, Steven",
    title = "{Damping of tensor modes in cosmology}",
    eprint = "astro-ph/0306304",
    archivePrefix = "arXiv",
    reportNumber = "UTTG-02-03",
    doi = "10.1103/PhysRevD.69.023503",
    journal = "Phys. Rev. D",
    volume = "69",
    pages = "023503",
    year = "2004"
}

@article{Anderson:1987yt,
    author = "Anderson, Paul R. and Parker, Leonard",
    title = "{Adiabatic Regularization in Closed Robertson-walker Universes}",
    reportNumber = "Print-87-0517 (MONTANA STATE)",
    doi = "10.1103/PhysRevD.36.2963",
    journal = "Phys. Rev. D",
    volume = "36",
    pages = "2963",
    year = "1987"
}

@book{Birrell:1982ix,
    author = "Birrell, N. D. and Davies, P. C. W.",
    title = "{Quantum Fields in Curved Space}",
    doi = "10.1017/CBO9780511622632",
    isbn = "978-0-511-62263-2, 978-0-521-27858-4",
    publisher = "Cambridge University Press",
    address = "Cambridge, UK",
    series = "Cambridge Monographs on Mathematical Physics",
    year = "1982"
}

@article{Parker:1974qw,
    author = "Parker, Leonard and Fulling, S. A.",
    title = "{Adiabatic regularization of the energy momentum tensor of a quantized field in homogeneous spaces}",
    doi = "10.1103/PhysRevD.9.341",
    journal = "Phys. Rev. D",
    volume = "9",
    pages = "341--354",
    year = "1974"
}

@article{Kofman:1997yn,
    author = "Kofman, Lev and Linde, Andrei D. and Starobinsky, Alexei A.",
    title = "{Towards the theory of reheating after inflation}",
    eprint = "hep-ph/9704452",
    archivePrefix = "arXiv",
    reportNumber = "IFA-97-28, SU-ITP-97-18",
    doi = "10.1103/PhysRevD.56.3258",
    journal = "Phys. Rev. D",
    volume = "56",
    pages = "3258--3295",
    year = "1997"
}

@article{Cook:2011hg,
    author = "Cook, Jessica L. and Sorbo, Lorenzo",
    title = "{Particle production during inflation and gravitational waves detectable by ground-based interferometers}",
    eprint = "1109.0022",
    archivePrefix = "arXiv",
    primaryClass = "astro-ph.CO",
    doi = "10.1103/PhysRevD.85.023534",
    journal = "Phys. Rev. D",
    volume = "85",
    pages = "023534",
    year = "2012",
    note = "[Erratum: Phys.Rev.D 86, 069901 (2012)]"
}

@article{Kolb:2023ydq,
    author = "Kolb, Edward W. and Long, Andrew J.",
    title = "{Cosmological gravitational particle production and its implications for cosmological relics}",
    eprint = "2312.09042",
    archivePrefix = "arXiv",
    primaryClass = "astro-ph.CO",
    doi = "10.1103/RevModPhys.96.045005",
    journal = "Rev. Mod. Phys.",
    volume = "96",
    number = "4",
    pages = "045005",
    year = "2024"
}

@article{Ema:2018ucl,
    author = "Ema, Yohei and Nakayama, Kazunori and Tang, Yong",
    title = "{Production of Purely Gravitational Dark Matter}",
    eprint = "1804.07471",
    archivePrefix = "arXiv",
    primaryClass = "hep-ph",
    reportNumber = "UT-18-08, KEK-TH-2047",
    doi = "10.1007/JHEP09(2018)135",
    journal = "JHEP",
    volume = "09",
    pages = "135",
    year = "2018"
}

@article{LIGOScientific:2022sts,
    author = "Abbott, R. and others",
    collaboration = "LIGO Scientific, KAGRA, VIRGO",
    title = "{Search for Gravitational-wave Transients Associated with Magnetar Bursts in Advanced LIGO and Advanced Virgo Data from the Third Observing Run}",
    eprint = "2210.10931",
    archivePrefix = "arXiv",
    primaryClass = "astro-ph.HE",
    reportNumber = "LIGO-P2100387",
    doi = "10.3847/1538-4357/ad27d3",
    journal = "Astrophys. J.",
    volume = "966",
    number = "1",
    pages = "137",
    year = "2024"
}

@book{CMB-S4:2016ple,
    author = "Abazajian, Kevork N. and others",
    collaboration = "CMB-S4",
    title = "{CMB-S4 Science Book, First Edition}",
    eprint = "1610.02743",
    archivePrefix = "arXiv",
    primaryClass = "astro-ph.CO",
    reportNumber = "FERMILAB-FN-1024-A-AE",
    doi = "10.2172/1352047",
    month = "10",
    year = "2016"
}

@article{Caravano:2022epk,
    author = "Caravano, Angelo and Komatsu, Eiichiro and Lozanov, Kaloian D. and Weller, Jochen",
    title = "{Lattice simulations of axion-U(1) inflation}",
    eprint = "2204.12874",
    archivePrefix = "arXiv",
    primaryClass = "astro-ph.CO",
    doi = "10.1103/PhysRevD.108.043504",
    journal = "Phys. Rev. D",
    volume = "108",
    number = "4",
    pages = "043504",
    year = "2023"
}

@article{Berghaus:2025dqi,
    author = "Berghaus, Kim V. and Drewes, Marco and Zell, Sebastian",
    title = "{Warm Inflation with the Standard Model}",
    eprint = "2503.18829",
    archivePrefix = "arXiv",
    primaryClass = "hep-ph",
    doi = "10.1103/9nn9-bsm9",
    journal = "Phys. Rev. Lett.",
    volume = "135",
    number = "17",
    pages = "171002",
    year = "2025"
}

@article{Ferreira:2017lnd,
    author = "Ferreira, Ricardo Z. and Notari, Alessio",
    title = "{Thermalized Axion Inflation}",
    eprint = "1706.00373",
    archivePrefix = "arXiv",
    primaryClass = "astro-ph.CO",
    doi = "10.1088/1475-7516/2017/09/007",
    journal = "JCAP",
    volume = "09",
    pages = "007",
    year = "2017"
}

@article{Martin:1997ns,
    author = "Martin, Stephen P.",
    editor = "Kane, Gordon L.",
    title = "{A Supersymmetry primer}",
    eprint = "hep-ph/9709356",
    archivePrefix = "arXiv",
    reportNumber = "FERMILAB-PUB-97-425-T",
    doi = "10.1142/9789812839657_0001",
    journal = "Adv. Ser. Direct. High Energy Phys.",
    volume = "18",
    pages = "1--98",
    year = "1998"
}

@article{An:2024zfi,
    author = "An, Haipeng and Chen, Qi and Yin, Yuan",
    title = "{A model for inflaton induced baryogenesis and its phenomenological consequences}",
    eprint = "2409.05833",
    archivePrefix = "arXiv",
    primaryClass = "astro-ph.CO",
    doi = "10.1007/JHEP06(2025)222",
    journal = "JHEP",
    volume = "06",
    pages = "222",
    year = "2025"
}

@article{An:2022toi,
    author = "An, Haipeng and Tong, Xi and Zhou, Siyi",
    title = "{Superheavy dark matter production from a symmetry-restoring first-order phase transition during inflation}",
    eprint = "2208.14857",
    archivePrefix = "arXiv",
    primaryClass = "hep-ph",
    reportNumber = "KOBE-COSMO-22-11",
    doi = "10.1103/PhysRevD.107.023522",
    journal = "Phys. Rev. D",
    volume = "107",
    number = "2",
    pages = "023522",
    year = "2023"
}

@article{Bodas:2020yho,
    author = "Bodas, Arushi and Kumar, Soubhik and Sundrum, Raman",
    title = "{The Scalar Chemical Potential in Cosmological Collider Physics}",
    eprint = "2010.04727",
    archivePrefix = "arXiv",
    primaryClass = "hep-ph",
    reportNumber = "UMD-PP-020-09",
    doi = "10.1007/JHEP02(2021)079",
    journal = "JHEP",
    volume = "02",
    pages = "079",
    year = "2021"
}

@article{Starobinsky:1979ty,
    author = "Starobinsky, Alexei A.",
    editor = "Khalatnikov, I. M. and Mineev, V. P.",
    title = "{Spectrum of relict gravitational radiation and the early state of the universe}",
    journal = "JETP Lett.",
    volume = "30",
    pages = "682--685",
    year = "1979"
}

@article{Guth:1980zm,
    author = "Guth, Alan H.",
    editor = "Fang, Li-Zhi and Ruffini, R.",
    title = "{The Inflationary Universe: A Possible Solution to the Horizon and Flatness Problems}",
    reportNumber = "SLAC-PUB-2576",
    doi = "10.1103/PhysRevD.23.347",
    journal = "Phys. Rev. D",
    volume = "23",
    pages = "347--356",
    year = "1981"
}

@article{Linde:1981mu,
    author = "Linde, Andrei D.",
    editor = "Fang, Li-Zhi and Ruffini, R.",
    title = "{A New Inflationary Universe Scenario: A Possible Solution of the Horizon, Flatness, Homogeneity, Isotropy and Primordial Monopole Problems}",
    reportNumber = "LEBEDEV-81-229",
    doi = "10.1016/0370-2693(82)91219-9",
    journal = "Phys. Lett. B",
    volume = "108",
    pages = "389--393",
    year = "1982"
}

@article{Albrecht:1982wi,
    author = "Albrecht, Andreas and Steinhardt, Paul J.",
    editor = "Fang, Li-Zhi and Ruffini, R.",
    title = "{Cosmology for Grand Unified Theories with Radiatively Induced Symmetry Breaking}",
    reportNumber = "UPR-0185T",
    doi = "10.1103/PhysRevLett.48.1220",
    journal = "Phys. Rev. Lett.",
    volume = "48",
    pages = "1220--1223",
    year = "1982"
}

@article{Affleck:1984fy,
    author = "Affleck, Ian and Dine, Michael",
    title = "{A New Mechanism for Baryogenesis}",
    reportNumber = "Print-84-0574 (PRINCETON)",
    doi = "10.1016/0550-3213(85)90021-5",
    journal = "Nucl. Phys. B",
    volume = "249",
    pages = "361--380",
    year = "1985"
}

@article{Mohapatra:2021aig,
    author = "Mohapatra, Rabindra N. and Okada, Nobuchika",
    title = "{Affleck-Dine baryogenesis with observable neutron-antineutron oscillation}",
    eprint = "2107.01514",
    archivePrefix = "arXiv",
    primaryClass = "hep-ph",
    doi = "10.1103/PhysRevD.104.055030",
    journal = "Phys. Rev. D",
    volume = "104",
    number = "5",
    pages = "055030",
    year = "2021"
}

@article{Hertzberg:2013mba,
    author = "Hertzberg, Mark P. and Karouby, Johanna",
    title = "{Generating the Observed Baryon Asymmetry from the Inflaton Field}",
    eprint = "1309.0010",
    archivePrefix = "arXiv",
    primaryClass = "hep-ph",
    reportNumber = "MIT-CTP-4493",
    doi = "10.1103/PhysRevD.89.063523",
    journal = "Phys. Rev. D",
    volume = "89",
    number = "6",
    pages = "063523",
    year = "2014"
}

@article{Lozanov:2014zfa,
    author = "Lozanov, Kaloian D. and Amin, Mustafa A.",
    title = "{End of inflation, oscillons, and matter-antimatter asymmetry}",
    eprint = "1408.1811",
    archivePrefix = "arXiv",
    primaryClass = "hep-ph",
    doi = "10.1103/PhysRevD.90.083528",
    journal = "Phys. Rev. D",
    volume = "90",
    number = "8",
    pages = "083528",
    year = "2014"
}

@article{Yamada:2015xyr,
    author = "Yamada, Masaki",
    title = "{Affleck-Dine baryogenesis just after inflation}",
    eprint = "1511.05974",
    archivePrefix = "arXiv",
    primaryClass = "hep-ph",
    reportNumber = "IPMU-15-195, DESY-15-215",
    doi = "10.1103/PhysRevD.93.083516",
    journal = "Phys. Rev. D",
    volume = "93",
    number = "8",
    pages = "083516",
    year = "2016"
}

@article{Bamba:2016vjs,
    author = "Bamba, Kazuharu and Barrie, Neil D. and Sugamoto, Akio and Takeuchi, Tatsu and Yamashita, Kimiko",
    title = "{Ratchet baryogenesis and an analogy with the forced pendulum}",
    eprint = "1610.03268",
    archivePrefix = "arXiv",
    primaryClass = "hep-ph",
    reportNumber = "OCHA-PP-342",
    doi = "10.1142/S0217732318500979",
    journal = "Mod. Phys. Lett. A",
    volume = "33",
    number = "17",
    pages = "1850097",
    year = "2018"
}

@article{Bamba:2018bwl,
    author = "Bamba, Kazuharu and Barrie, Neil D. and Sugamoto, Akio and Takeuchi, Tatsu and Yamashita, Kimiko",
    title = "{Pendulum Leptogenesis}",
    eprint = "1805.04826",
    archivePrefix = "arXiv",
    primaryClass = "hep-ph",
    reportNumber = "OCHA-PP-351",
    doi = "10.1016/j.physletb.2018.08.044",
    journal = "Phys. Lett. B",
    volume = "785",
    pages = "184--190",
    year = "2018"
}

@article{Barrie:2020hiu,
    author = "Barrie, Neil D. and Sugamoto, Akio and Takeuchi, Tatsu and Yamashita, Kimiko",
    title = "{Higgs Inflation, Vacuum Stability, and Leptogenesis}",
    eprint = "2001.07032",
    archivePrefix = "arXiv",
    primaryClass = "hep-ph",
    reportNumber = "IPMU19-0175, OCHA-PP-359",
    doi = "10.1007/JHEP08(2020)072",
    journal = "JHEP",
    volume = "08",
    pages = "072",
    year = "2020"
}

@article{Lin:2020lmr,
    author = "Lin, Chia-Min and Kohri, Kazunori",
    title = "{Inflaton as the Affleck-Dine Baryogenesis Field in Hilltop Supernatural Inflation}",
    eprint = "2003.13963",
    archivePrefix = "arXiv",
    primaryClass = "hep-ph",
    reportNumber = "KEK-Cosmo-250, KEK-TH-2203, IPMU20-0037",
    doi = "10.1103/PhysRevD.102.043511",
    journal = "Phys. Rev. D",
    volume = "102",
    number = "4",
    pages = "043511",
    year = "2020"
}

@article{Wu:2019ohx,
    author = "Wu, Yi-Peng and Yang, Louis and Kusenko, Alexander",
    title = "{Leptogenesis from spontaneous symmetry breaking during inflation}",
    eprint = "1905.10537",
    archivePrefix = "arXiv",
    primaryClass = "hep-ph",
    reportNumber = "RESCEU 6/19, IPMU19-0079",
    doi = "10.1007/JHEP12(2019)088",
    journal = "JHEP",
    volume = "12",
    pages = "088",
    year = "2019"
}

@article{Charng:2008ke,
    author = "Charng, Yeo-Yie and Lee, Da-Shin and Leung, Chung Ngoc and Ng, Kin-Wang",
    title = "{Affleck-Dine Baryogenesis, Split Supersymmetry, and Inflation}",
    eprint = "0802.1328",
    archivePrefix = "arXiv",
    primaryClass = "hep-ph",
    doi = "10.1103/PhysRevD.80.063519",
    journal = "Phys. Rev. D",
    volume = "80",
    pages = "063519",
    year = "2009"
}

@article{NANOGrav:2023hde,
            author = "Agazie, Gabriella and others",
            collaboration = "NANOGrav",
            title = "{The NANOGrav 15 yr Data Set: Observations and Timing of 68 Millisecond Pulsars}",
            eprint = "2306.16217",
            archivePrefix = "arXiv",
            primaryClass = "astro-ph.HE",
            doi = "10.3847/2041-8213/acda9a",
            journal = "Astrophys. J. Lett.",
            volume = "951",
            number = "1",
            pages = "L9",
            year = "2023"
        }

@article{EPTA:2023sfo,
    author = "Antoniadis, J. and others",
    collaboration = "EPTA",
    title = "{The second data release from the European Pulsar Timing Array - I. The dataset and timing analysis}",
    eprint = "2306.16224",
    archivePrefix = "arXiv",
    primaryClass = "astro-ph.HE",
    doi = "10.1051/0004-6361/202346841",
    journal = "Astron. Astrophys.",
    volume = "678",
    pages = "A48",
    year = "2023"
}

@article{Audley:2017drz,
            author = "Amaro-Seoane, Pau and others",
            collaboration = "LISA",
            title = "{Laser Interferometer Space Antenna}",
            eprint = "1702.00786",
            archivePrefix = "arXiv",
            primaryClass = "astro-ph.IM",
            month = "2",
            year = "2017"
        }

@article{Luo:2015ght,
            author = "Luo, Jun and others",
            collaboration = "TianQin",
            title = "{TianQin: a space-borne gravitational wave detector}",
            eprint = "1512.02076",
            archivePrefix = "arXiv",
            primaryClass = "astro-ph.IM",
            doi = "10.1088/0264-9381/33/3/035010",
            journal = "Class. Quant. Grav.",
            volume = "33",
            number = "3",
            pages = "035010",
            year = "2016"
        }

@article{Guo:2018npi,
            author = "Ruan, Wen-Hong and Guo, Zong-Kuan and Cai, Rong-Gen and Zhang, Yuan-Zhong",
            title = "{Taiji Program: Gravitational-Wave Sources}",
            eprint = "1807.09495",
            archivePrefix = "arXiv",
            primaryClass = "gr-qc",
            doi = "10.1142/S0217751X2050075X",
            journal = "Int. J. Mod. Phys. A",
            volume = "35",
            number = "17",
            pages = "2050075",
            year = "2020"
        }

@article{Berera:1995ie,
            author = "Berera, Arjun",
            title = "{Warm inflation}",
            eprint = "astro-ph/9509049",
            archivePrefix = "arXiv",
            reportNumber = "PSU-TH-159",
            doi = "10.1103/PhysRevLett.75.3218",
            journal = "Phys. Rev. Lett.",
            volume = "75",
            pages = "3218--3221",
            year = "1995"
        }

@Misc{ArkaniHamed2015,
  author        = {Nima Arkani-Hamed and Juan Maldacena},
  title         = {Cosmological Collider Physics},
  year          = {2015},
  archiveprefix = {arXiv},
  eprint        = {1503.08043},
  primaryclass  = {hep-th},
}

@article{Aoki:2023tjm,
    author = "Aoki, Shuntaro",
    title = "{Continuous spectrum on cosmological collider}",
    eprint = "2301.07920",
    archivePrefix = "arXiv",
    primaryClass = "hep-th",
    doi = "10.1088/1475-7516/2023/04/002",
    journal = "JCAP",
    volume = "04",
    pages = "002",
    year = "2023"
}

@article{Aoki:2025uff,
    author = "Aoki, Shuntaro and Strumia, Alessandro",
    title = "{Testing the arrow of time at the cosmo collider}",
    eprint = "2510.05204",
    archivePrefix = "arXiv",
    primaryClass = "hep-ph",
    reportNumber = "RIKEN-iTHEMS-Report-25",
    month = "10",
    year = "2025"
}

@article{Aoki:2024jha,
    author = "Aoki, Shuntaro and Ghoshal, Anish and Strumia, Alessandro",
    title = "{Cosmological collider non-Gaussianity from multiple scalars and R$^{2}$ gravity}",
    eprint = "2408.07069",
    archivePrefix = "arXiv",
    primaryClass = "astro-ph.CO",
    doi = "10.1007/JHEP11(2024)009",
    journal = "JHEP",
    volume = "11",
    pages = "009",
    year = "2024"
}

@Article{Chen2017,
  author        = {Chen, Xingang and Wang, Yi and Xianyu, Zhong-Zhi},
  journal       = {JCAP},
  title         = {{Schwinger-Keldysh Diagrammatics for Primordial Perturbations}},
  year          = {2017},
  pages         = {006},
  volume        = {12},
  archiveprefix = {arXiv},
  doi           = {10.1088/1475-7516/2017/12/006},
  eprint        = {1703.10166},
  primaryclass  = {hep-th},
}

@Article{Wang2020,
  author        = {Wang, Lian-Tao and Xianyu, Zhong-Zhi},
  journal       = {JHEP},
  title         = {{In Search of Large Signals at the Cosmological Collider}},
  year          = {2020},
  pages         = {044},
  volume        = {02},
  archiveprefix = {arXiv},
  doi           = {10.1007/JHEP02(2020)044},
  eprint        = {1910.12876},
  primaryclass  = {hep-ph},
}

@article{Tong:2022cdz,
    author = "Tong, Xi and Xianyu, Zhong-Zhi",
    title = "{Large spin-2 signals at the cosmological collider}",
    eprint = "2203.06349",
    archivePrefix = "arXiv",
    primaryClass = "hep-ph",
    doi = "10.1007/JHEP10(2022)194",
    journal = "JHEP",
    volume = "10",
    pages = "194",
    year = "2022"
}

@article{Yin:2023jlv,
    author = "Yin, Yuan",
    title = "{Cosmological collider signal from non-Bunch-Davies initial states}",
    eprint = "2309.05244",
    archivePrefix = "arXiv",
    primaryClass = "hep-ph",
    doi = "10.1103/PhysRevD.109.043535",
    journal = "Phys. Rev. D",
    volume = "109",
    number = "4",
    pages = "043535",
    year = "2024"
}

@article{Hubisz:2024xnj,
    author = "Hubisz, Jay and Lee, Seung J. and Li, He and Sambasivam, Bharath",
    title = "{Cosmological quasiparticles and the cosmological collider}",
    eprint = "2408.08951",
    archivePrefix = "arXiv",
    primaryClass = "astro-ph.CO",
    doi = "10.1103/PhysRevD.111.023543",
    journal = "Phys. Rev. D",
    volume = "111",
    number = "2",
    pages = "023543",
    year = "2025"
}

@Article{Pimentel2022,
  author        = {Pimentel, Guilherme L. and Wang, Dong-Gang},
  journal       = {JHEP},
  title         = {{Boostless cosmological collider bootstrap}},
  year          = {2022},
  pages         = {177},
  volume        = {10},
  archiveprefix = {arXiv},
  doi           = {10.1007/JHEP10(2022)177},
  eprint        = {2205.00013},
  primaryclass  = {hep-th},
}

@Article{Chen2010,
  author        = {Chen, Xingang and Wang, Yi},
  journal       = {JCAP},
  title         = {{Quasi-Single Field Inflation and Non-Gaussianities}},
  year          = {2010},
  pages         = {027},
  volume        = {04},
  archiveprefix = {arXiv},
  doi           = {10.1088/1475-7516/2010/04/027},
  eprint        = {0911.3380},
  primaryclass  = {hep-th},
}

@Article{Chen2018,
  author        = {Chen, Xingang and Wang, Yi and Xianyu, Zhong-Zhi},
  journal       = {JHEP},
  title         = {{Neutrino Signatures in Primordial Non-Gaussianities}},
  year          = {2018},
  pages         = {022},
  volume        = {09},
  archiveprefix = {arXiv},
  doi           = {10.1007/JHEP09(2018)022},
  eprint        = {1805.02656},
  primaryclass  = {hep-ph},
}

@Article{Wang2020a,
  author        = {Wang, Lian-Tao and Xianyu, Zhong-Zhi},
  journal       = {JHEP},
  title         = {{Gauge Boson Signals at the Cosmological Collider}},
  year          = {2020},
  pages         = {082},
  volume        = {11},
  archiveprefix = {arXiv},
  doi           = {10.1007/JHEP11(2020)082},
  eprint        = {2004.02887},
  primaryclass  = {hep-ph},
}

@Article{Cui2022,
  author        = {Cui, Yanou and Xianyu, Zhong-Zhi},
  journal       = {Phys. Rev. Lett.},
  title         = {{Probing Leptogenesis with the Cosmological Collider}},
  year          = {2022},
  number        = {11},
  pages         = {111301},
  volume        = {129},
  archiveprefix = {arXiv},
  doi           = {10.1103/PhysRevLett.129.111301},
  eprint        = {2112.10793},
  primaryclass  = {hep-ph},
}

@Article{Lee2016,
  author        = {Lee, Hayden and Baumann, Daniel and Pimentel, Guilherme L.},
  journal       = {JHEP},
  title         = {{Non-Gaussianity as a Particle Detector}},
  year          = {2016},
  pages         = {040},
  volume        = {12},
  archiveprefix = {arXiv},
  doi           = {10.1007/JHEP12(2016)040},
  eprint        = {1607.03735},
  primaryclass  = {hep-th},
}

@article{Chen:2015lza,
    author = "Chen, Xingang and Namjoo, Mohammad Hossein and Wang, Yi",
    title = "{Quantum Primordial Standard Clocks}",
    eprint = "1509.03930",
    archivePrefix = "arXiv",
    primaryClass = "astro-ph.CO",
    doi = "10.1088/1475-7516/2016/02/013",
    journal = "JCAP",
    volume = "02",
    pages = "013",
    year = "2016"
}

@article{Chen:2016nrs,
    author = "Chen, Xingang and Wang, Yi and Xianyu, Zhong-Zhi",
    title = "{Loop Corrections to Standard Model Fields in Inflation}",
    eprint = "1604.07841",
    archivePrefix = "arXiv",
    primaryClass = "hep-th",
    doi = "10.1007/JHEP08(2016)051",
    journal = "JHEP",
    volume = "08",
    pages = "051",
    year = "2016"
}

@article{Chen:2016uwp,
    author = "Chen, Xingang and Wang, Yi and Xianyu, Zhong-Zhi",
    title = "{Standard Model Background of the Cosmological Collider}",
    eprint = "1610.06597",
    archivePrefix = "arXiv",
    primaryClass = "hep-th",
    doi = "10.1103/PhysRevLett.118.261302",
    journal = "Phys. Rev. Lett.",
    volume = "118",
    number = "26",
    pages = "261302",
    year = "2017"
}

@article{Chen:2016hrz,
    author = "Chen, Xingang and Wang, Yi and Xianyu, Zhong-Zhi",
    title = "{Standard Model Mass Spectrum in Inflationary Universe}",
    eprint = "1612.08122",
    archivePrefix = "arXiv",
    primaryClass = "hep-th",
    doi = "10.1007/JHEP04(2017)058",
    journal = "JHEP",
    volume = "04",
    pages = "058",
    year = "2017"
}

@article{Ema:2015dka,
    author = "Ema, Yohei and Jinno, Ryusuke and Mukaida, Kyohei and Nakayama, Kazunori",
    title = "{Gravitational Effects on Inflaton Decay}",
    eprint = "1502.02475",
    archivePrefix = "arXiv",
    primaryClass = "hep-ph",
    reportNumber = "UT-15-03",
    doi = "10.1088/1475-7516/2015/05/038",
    journal = "JCAP",
    volume = "05",
    pages = "038",
    year = "2015"
}

@article{Ema:2016hlw,
    author = "Ema, Yohei and Jinno, Ryusuke and Mukaida, Kyohei and Nakayama, Kazunori",
    title = "{Gravitational particle production in oscillating backgrounds and its cosmological implications}",
    eprint = "1604.08898",
    archivePrefix = "arXiv",
    primaryClass = "hep-ph",
    reportNumber = "UT-16-20, KEK-TH-1898, IPMU-16-0062",
    doi = "10.1103/PhysRevD.94.063517",
    journal = "Phys. Rev. D",
    volume = "94",
    number = "6",
    pages = "063517",
    year = "2016"
}

@article{Li:2019ves,
    author = "Li, Lingfeng and Nakama, Tomohiro and Sou, Chon Man and Wang, Yi and Zhou, Siyi",
    title = "{Gravitational Production of Superheavy Dark Matter and Associated Cosmological Signatures}",
    eprint = "1903.08842",
    archivePrefix = "arXiv",
    primaryClass = "astro-ph.CO",
    doi = "10.1007/JHEP07(2019)067",
    journal = "JHEP",
    volume = "07",
    pages = "067",
    year = "2019"
}

@article{Li:2020xwr,
    author = "Li, Lingfeng and Lu, Shiyun and Wang, Yi and Zhou, Siyi",
    title = "{Cosmological Signatures of Superheavy Dark Matter}",
    eprint = "2002.01131",
    archivePrefix = "arXiv",
    primaryClass = "hep-ph",
    doi = "10.1007/JHEP07(2020)231",
    journal = "JHEP",
    volume = "07",
    pages = "231",
    year = "2020"
}

@article{Aoki:2020zbj,
    author = "Aoki, Shuntaro and Yamaguchi, Masahide",
    title = "{Disentangling mass spectra of multiple fields in cosmological collider}",
    eprint = "2012.13667",
    archivePrefix = "arXiv",
    primaryClass = "hep-th",
    reportNumber = "WU-HEP-20-15",
    doi = "10.1007/JHEP04(2021)127",
    journal = "JHEP",
    volume = "04",
    pages = "127",
    year = "2021"
}

@article{Pinol:2021aun,
    author = "Pinol, Lucas and Aoki, Shuntaro and Renaux-Petel, S\'ebastien and Yamaguchi, Masahide",
    title = "{Inflationary flavor oscillations and the cosmic spectroscopy}",
    eprint = "2112.05710",
    archivePrefix = "arXiv",
    primaryClass = "hep-th",
    doi = "10.1103/PhysRevD.107.L021301",
    journal = "Phys. Rev. D",
    volume = "107",
    number = "2",
    pages = "L021301",
    year = "2023"
}

@article{LISA:2024mow,
    author = "Colpi, Monica and others",
    collaboration = "LISA",
    title = "{LISA Definition Study Report}",
    eprint = "2402.07571",
    archivePrefix = "arXiv",
    primaryClass = "astro-ph.CO",
    month = "2",
    year = "2024"
}

@article{Robson:2018ifk,
    author = "Robson, Travis and Cornish, Neil J. and Liu, Chang",
    title = "{The construction and use of LISA sensitivity curves}",
    eprint = "1803.01944",
    archivePrefix = "arXiv",
    primaryClass = "astro-ph.HE",
    doi = "10.1088/1361-6382/ab1101",
    journal = "Class. Quant. Grav.",
    volume = "36",
    number = "10",
    pages = "105011",
    year = "2019"
}

@article{Babak:2023lro,
    author = "Babak, Stanislav and Caprini, Chiara and Figueroa, Daniel G. and Karnesis, Nikolaos and Marcoccia, Paolo and Nardini, Germano and Pieroni, Mauro and Ricciardone, Angelo and Sesana, Alberto and Torrado, Jes\'us",
    title = "{Stochastic gravitational wave background from stellar origin binary black holes in LISA}",
    eprint = "2304.06368",
    archivePrefix = "arXiv",
    primaryClass = "astro-ph.CO",
    month = "4",
    year = "2023"
}

@article{Tristram:2021tvh,
    author = "Tristram, M. and others",
    title = "{Improved limits on the tensor-to-scalar ratio using BICEP and Planck}",
    eprint = "2112.07961",
    archivePrefix = "arXiv",
    primaryClass = "astro-ph.CO",
    doi = "10.1103/PhysRevD.105.083524",
    journal = "Phys. Rev. D",
    volume = "105",
    number = "8",
    pages = "083524",
    year = "2022"
}
\bibliographystyle{utphys}

\end{document}